\documentstyle[twocolumn]{mn}
\newif\ifAMStwofonts
\ifoldfss
  \ifCUPmtlplainloaded \else
    \NewTextAlphabet{textbfit} {cmbxti10} {}
    \NewTextAlphabet{textbfss} {cmssbx10} {}
    \NewMathAlphabet{mathbfit} {cmbxti10} {} 
    \NewMathAlphabet{mathbfss} {cmssbx10} {} 
  \fi
  \ifAMStwofonts
    \ifCUPmtlplainloaded \else
      \NewSymbolFont{upmath} {eurm10}
      \NewSymbolFont{AMSa} {msam10}
      \NewMathSymbol{\upi}     {0}{upmath}{19}
      \NewMathSymbol{\umu}     {0}{upmath}{16}
      \NewMathSymbol{\upartial}{0}{upmath}{40}
      \NewMathSymbol{\leqslant}{3}{AMSa}{36}
      \NewMathSymbol{\geqslant}{3}{AMSa}{3E}

       \let\le=\leqslant
       \let\ge=\geqslant
    \fi
  \fi
\fi 

\ifnfssone
  \newmathalphabet{\mathit}
  \addtoversion{normal}{\mathit}{cmr}{m}{it}
  \addtoversion{bold}{\mathit}{cmr}{bx}{it}
  \newmathalphabet{\mathbfit} 
  \addtoversion{normal}{\mathbfit}{cmr}{bx}{it}
  \addtoversion{bold}{\mathbfit}{cmr}{bx}{it}
  \newmathalphabet{\mathbfss} 
  \addtoversion{normal}{\mathbfss}{cmss}{bx}{n}
  \addtoversion{bold}{\mathbfss}{cmss}{bx}{n}
  \ifAMStwofonts
    \ifCUPmtlplainloaded \else
      \UseAMStwoboldmath
      \makeatletter
      \new@mathgroup\upmath@group
      \define@mathgroup\mv@normal\upmath@group{eur}{m}{n}
      \define@mathgroup\mv@bold\upmath@group{eur}{b}{n}
      \edef\UPM{\hexnumber\upmath@group}
      \new@mathgroup\amsa@group
      \define@mathgroup\mv@normal\amsa@group{msa}{m}{n}
      \define@mathgroup\mv@bold\amsa@group{msa}{m}{n}
      \edef\AMSa{\hexnumber\amsa@group}
      \makeatother
      \mathchardef\upi="0\UPM19
      \mathchardef\umu="0\UPM16
      \mathchardef\upartial="0\UPM40
      \mathchardef\leqslant="3\AMSa36
      \mathchardef\geqslant="3\AMSa3E

       \let\le=\leqslant
       \let\ge=\geqslant
    \fi
  \fi
\fi 

\ifnfsstwo
  \DeclareMathAlphabet{\mathbfit}{OT1}{cmr}{bx}{it}
  \SetMathAlphabet\mathbfit{bold}{OT1}{cmr}{bx}{it}
  \DeclareMathAlphabet{\mathbfss}{OT1}{cmss}{bx}{n}
  \SetMathAlphabet\mathbfss{bold}{OT1}{cmss}{bx}{n}
  \ifAMStwofonts
    \ifCUPmtlplainloaded \else
      \DeclareSymbolFont{UPM}{U}{eur}{m}{n}
      \SetSymbolFont{UPM}{bold}{U}{eur}{b}{n}
      \DeclareSymbolFont{AMSa}{U}{msa}{m}{n}
      \DeclareMathSymbol{\upi}{0}{UPM}{"19}
      \DeclareMathSymbol{\umu}{0}{UPM}{"16}
      \DeclareMathSymbol{\upartial}{0}{UPM}{"40}
      \DeclareMathSymbol{\leqslant}{3}{AMSa}{"36}
      \DeclareMathSymbol{\geqslant}{3}{AMSa}{"3E}

       \let\le=\leqslant
       \let\ge=\geqslant
    \fi
  \fi
\fi 

\ifCUPmtlplainloaded \else
  \ifAMStwofonts \else 
    \def\upi{\pi}
    \def\umu{\mu}
    \def\upartial{\partial}
  \fi
\fi

\title[The internal dynamics of the Local Group dEs]{The internal
dynamics of the Local Group dwarf elliptical galaxies NGC147, NGC185,
and NGC205 \thanks{Based on observations collected at the Observatoire
de Haute-Provence.}}  \author[S. De Rijcke, P. Prugniel, F. Simien,
H. Dejonghe] {S. De Rijcke$^1$ \thanks{corresponding author:
sven.derijcke@UGent.be}\thanks{Postdoctoral Fellow of the Fund for
Scientific Research - Flanders (Belgium)(F.W.O)}, P. Prugniel$^2$,
F. Simien$^2$, H. Dejonghe$^1$ \\ $^1$ Sterrenkundig Observatorium,
Universiteit Gent, Krijgslaan 281, S9, B-9000, Gent, Belgium\\ $^2$
CRAL-Observatoire de Lyon, 9 Av. C. Andr\'e, 69561 Saint-Genis Laval,
France} \date{Accepted 1988 December 15.  Received 1988 December 14;
in original form 1988 October 11}

\pagerange{\pageref{firstpage}--\pageref{lastpage}}
\pubyear{1994}

\begin{document}

\maketitle

\label{firstpage}

\begin{abstract}
We present three-integral dynamical models for the three Local Group
dwarf elliptical galaxies:~NGC147, NGC185, and NGC205. These models
are fitted to the 2MASS J-band surface brightness distribution and the
major-axis kinematics (mean streaming velocity and velocity
dispersion) and, in the case of NGC205, also to the minor-axis
kinematics. The kinematical information extends out to 2~$R_{\rm e}$
in the case of NGC205 and out to about 1~$R_{\rm e}$ in the case of
NGC147 and NGC185. It is the first time models are constructed for the
Local Group dEs that allow for the presence of dark matter at large
radii and that are constrained by kinematics out to at least one
half-light radius. The B-band mass-to-light ratios of all three
galaxies are rather similar, $(M/L)_{\rm B} \approx 3-4
\,M_\odot/L_{\odot,\rm B}$. Within the inner two half-light radii,
about $40-50$\% of the mass is in the form of dark matter, so dEs
contain about as much dark matter as bright ellipticals.

Based on their appreciable apparent flattening, we modeled NGC205 and
NGC147 as being viewed edge-on. For NGC185, having a much rounder
appearance on the sky, we produced models for different inclinations.
NGC205 and NGC147 have a relatively isotropic velocity dispersion
tensor within the region where the internal dynamics are strongly
constrained by the data. Our estimated inclination for NGC185 is $i
\approx 50^\circ$ because in that case the model has an intrinsic
flattening close to the peak of the intrinsic shape distribution of
dEs and it, like the best fitting models for NGC147 and NGC205, is
nearly isotropic. We also show that the dynamical properties of the
bright nucleus of NGC205 are not unlike those of a massive globular
cluster.
\end{abstract}


\section{Introduction}

From the days of the great comet-hunter Charles Messier up to well
into the 20th century, NGC205 was the only known member of the class
of objects that we now call dwarf ellipticals (dEs) \cite{m01}. dEs
are faint ($M_B \ge -18$~mag) galaxies with diffuse, approximately
exponentially declining surface brightness profiles and smooth
elliptical isophotes \cite{fb94}. NGC205 was resolved into stars and
thus confirmed as a Local-Group member by Walter Baade, using the
100~inch telescope at Mount Wilson Observatory \cite{b44a}. He
correctly identified NGC205, at a projected distance of only
37{\arcmin} from M31, as a diffuse stellar system with a stellar
population very much like that of the Galactic globular
clusters. NGC147 and NGC185 form a close pair, separated by only
58{\arcmin} on the sky \cite{s98} and located some 7$^\circ$ away from
M31. Baade confirmed NGC147 and NGC185 as members of the Local Group
by resolving them into individual stars \cite{b44b}. No new Local
Group dEs have been discovered since then while the sensus of the
Local Group dwarf spheroidals (dSph), which are even fainter (absolute
B-band magnitude $M_B \ge -13$~mag) and more diffuse than dEs, is most
likely not yet complete. Since Baade could resolve these dwarf
galaxies at about the same magnitude as M31, he surmised that they
should be at roughly the same distance as M31. More precise distance
measurements have since then clarified the geometry of the M31
satellite galaxy system. Using the TRGB distances measured by
McConnachie \shortcite{mc05}, $D_{\rm NGC147} = 675 \pm 27$~kpc,
$D_{\rm NGC185} = 616 \pm 26$~kpc, $D_{\rm NGC205} = 824 \pm 27$~kpc,
$D_{\rm M31} = 785 \pm 25$~kpc, which are in good agreement with those
presented by Lee et al. \shortcite{l1} and Lee \shortcite{l2}, and the
angular distances between the galaxies, the following picture
emerges. Using the distance moduli and the sources of photometric
errors given by McConnachie et al. \shortcite{mc05}, we find that
NGC147 and NGC185 are separated by $63 \pm 33$~kpc and the couple is
situated 160~kpc in front of M31. For all practical purposes, we can
therefore treat them as isolated stellar systems. NGC205, on the other
hand, is situated $48 \pm 30$~kpc behind M31.

\begin{figure}
\vspace*{8cm}
\special{hscale=60 vscale=60 hsize=500 vsize=240
hoffset=-20 voffset=-60 angle=0 psfile="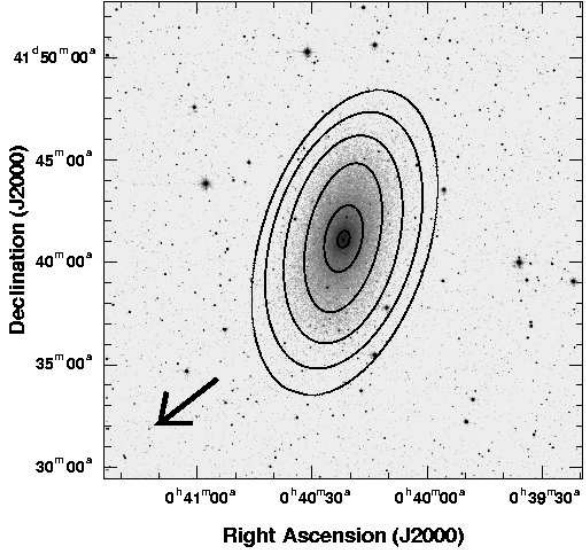"}
\caption{2MASS J-band image of NGC205. Overplotted onto this image are
the fitted elliptical isophotes at J-band surface brightness levels of
18, 19, 20, 21, 22, and 23~mag~arcsec$^{-2}$. Beyond a major-axis
distance of $\approx 300${\arcsec} the isophotes of NGC205 start to
twist, by almost 10$^\circ$ between the 21 to 23 mag~arcsec$^{-2}$
isophotes. The direction towards the center of M31 is indicated by an
arrow.
\label{phot205}}
\end{figure}

In this paper, we present dynamical models for NGC147, NGC185, and
NGC205. In order to construct a mass model for the luminous matter, we
use 2MASS J-band images of these galaxies, downloaded from the
NASA/IPAC Infrared Science Archive website. These dwarf galaxies have
complex star-formation histories and show evidence for spatial
variations of the stellar population, the most striking being the very
blue nucleus of NGC205 \cite{bm05,ha97,p93}. Therefore, the density
distribution obtained from near-infrared J-band images should trace
the mass distribution of the luminous matter more closely than that
obtained from optical images. We measured the surface-brightness
profile, position angle, and ellipticity $\epsilon = 1-b/a$ of all
three galaxies as a function of the geometric mean of major and minor
axis distance, denoted by $a$ and $b$ respectively. These were
obtained using our own software. Basically, the code fits an ellipse
through a set of positions where a given surface brightness level is
reached. The shape of an isophote, relative to the best fitting
ellipse with semi-major axis a and ellipticity $\epsilon$, is
quantified by expanding the intensity variation along this ellipse in
a fourth order Fourier series with coefficients $S_4$, $S_3$, $C_4$
and $C_3$:
\begin{eqnarray}
 I(a,\theta) &=& I_0(a) \left[ 1 + C_3(a) \cos(3\theta)+
\right. \nonumber \\ && \hspace{-4em} \left. C_4(a) \cos(4\theta)) +
S_3(a)\sin(3\theta))+ S_4(a) \sin(4\theta) \right].
\end{eqnarray}
Here, $I_0(a)$ is the average intensity of the isophote and the angle
$\theta$ is measured from the major axis. Using all photometric
parameters, evaluated as a function of $a$, a smooth image of each
galaxy is reconstructed. We used this smooth image to measure the
total J-band magnitude and half-light radii. In each case, the total
apparent magnitude we derive is within 0.05~mag of that listed in the
2MASS Large Galaxy Atlas \cite{ja03}. As to the half-light radii, we
find for NGC205 that $R_{\rm e}({\rm NGC205}) = 130${\arcsec}
(0.52~kpc), for NGC147 that $R_{\rm e}({\rm NGC147}) = 122${\arcsec}
(0.38~kpc), and NGC185 that $R_{\rm e}({\rm NGC185}) = 90${\arcsec}
(0.27~kpc). 

This smooth image is then deprojected in order to obtain the spatial
density distribution of the stars. The spatial density of an
axisymmetric stellar system cannot uniquely be reconstructed from the
observed projected density if it is not viewed edge-on
\cite{gb96}. From the infinity of equally plausible positive, axially
symmetric spatial densities that project to the observed projected
density, we pick the one that can be represented as a weighted sum of
basis functions, $\rho(\varpi,z) = \sum_k c_k \, \rho_k(\varpi,z)$,
with $\rho_k(\varpi,z)$ of the form
\begin{equation}
\rho_k(\varpi,z) = \exp\left[
-\left(\frac{\sqrt{\varpi^2+(z/q_k)^2}}{\sigma_k} \right)^{m_k} \right].
\end{equation}
Here, $\sigma_k$ is a scale radius, $m_k$ is a shape parameter, making
the density profile more (small $m_k$) or less (large $m_k$) centrally
concentrated, $q$ is the axis-ratio, controlling the flattening of the
basis function. These basis functions produce elliptical isophotes
when projected onto the sky. For a given inclination, the unknown
coefficients in the expansion of the spatial density are determined by
fitting the projected basis functions to the observed photometry,
evaluated on an elliptical, logarithmically spaced grid subject to the
constraint that the spatial density is positive everywhere using a
Quadratic Programming algorithm \cite{d89}. Note that the coefficients
$c_k$ can be negative, as long as the weighted sum of basis functions
is positive. This algorithm has a library at its disposal that
contains hundreds of basis functions to choose from. Typically, the
$\chi^2$ does not further improve after co-adding a few tens of
components and the procedure can be stopped. This deprojection
procedure ensures that the spatial density is smooth, well-behaved,
and positive.

\begin{figure}
\vspace*{9.75cm}
\special{hscale=60 vscale=60 hsize=500 vsize=280
hoffset=-20 voffset=-60 angle=0 psfile="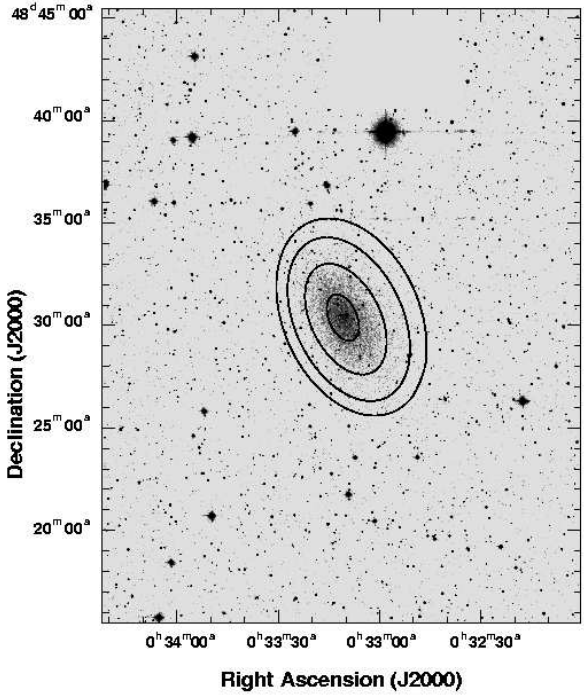"}
\caption{2MASS J-band image of NGC147. Overplotted onto this image are
the fitted elliptical isophotes at J-band surface brightness levels of
20, 21, 22, and 23~mag~arcsec$^{-2}$. \label{phot147}}
\end{figure}

\begin{figure}
\vspace*{8cm}
\special{hscale=60 vscale=60 hsize=500 vsize=240
hoffset=-20 voffset=-60 angle=0 psfile="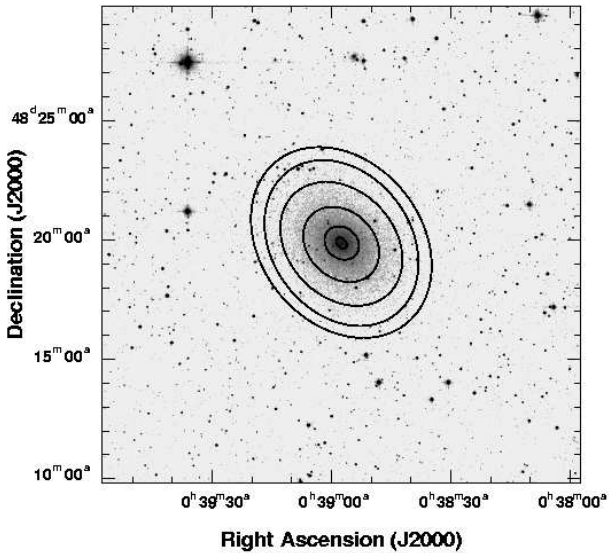"}
\caption{2MASS J-band image of NGC185. Overplotted onto this image are
the fitted elliptical isophotes at J-band surface brightness levels of
18, 19, 20, 21, 22, and 23~mag~arcsec$^{-2}$. \label{phot185}}
\end{figure}

Already from the 2MASS image, it is obvious that NGC205 is noticeably
influenced by the tidal forces exerted by M31 (see
Fig. \ref{phot205}). Beyond a major-axis distance of $\approx
300${\arcsec} the isophotes of NGC205 twist from a position angle of
167$^\circ$ to 160$^\circ$ at 400{\arcsec}. At that radius, the
surface brightness profile also shows a downward break, as can clearly
be seen in Fig. \ref{kin205} (see also Hodge \shortcite{h73} and Choi
et al. \shortcite{ch02}). In these outer regions, peculiar kinematics,
such as the outermost stars moving counter to the rotation of the main
body \cite{g04}, provide further evidence for a tidal interaction with
M31. We will therefore check if our results for NGC205 are affected by
it being so close to M31. The isophotes of NGC147 and NGC185 are much
more regular, with negligible isophote twisting, see
Figs. \ref{phot147} and \ref{phot185}.

The spectra were obtained with the 1.93-m telescope of the
Observatoire de Haute-Provence, equipped with the CARELEC long-slit
spectrograph. In combination with a EEV receptor with $2048\times780$
pixels of $15~\mu$m, the instrumental velocity dispersion was $\simeq
25$~km~s$^{-1}$ (R=5050) for a slit width of 1.5{\arcsec}. More
details and a log of the observations can be found in Simien \&
Prugniel \shortcite{sp02}. 10-minute exposures on blank fields of sky
obtained during the same night as the science exposures were used for
sky subtraction. With a Fourier-Fitting technique \cite{f89}, the
spectrum of a G8{\sc iii} template star (HD5395), convolved with a
Gaussian line-of-sight velocity distribution, yielded a good fit to
the galaxy spectrum in Fourier space, providing a simultaneous
estimate of the mean streaming velocity $v_p$ and the projected
velocity dispersion $\sigma_p$. 

From the deep major axis spectrum of NGC205, stellar kinematics out to
about 300{\arcsec} (1.2~kpc or 2.3 half-light radii) could be
extracted. The minor axis spectra yields useful kinematic information
out to 120{\arcsec} (0.9~kpc). Thanks to the proximity of this galaxy
and the unprecedented depth of the spectra, we can measure kinematics
near the center of NGC205 with a spatial resolution of $\sim
5-10$~parsecs. The major axis spectrum of NGC185 allows extracting
kinematics out to $\sim 100${\arcsec} (0.3~kpc or 1.1~$R_{\rm
e}$). NGC147 has the lowest surface brightness of all Local Group dEs
and is moreover riddled with foreground stars and was therefore the
most difficult object to obtain decent kinematics for. The spectra of
NGC147 secured by Simien \& Prugniel \shortcite{sp02} turned out to be
insufficiently deep. We, therefore, re-observed NGC147 in January 2002
and January 2003, using the same instrumental setup and the same data
reduction procedure. With the addition of these new exposures (six
one-hour exposures along the southern semi-major axis and four
one-hour exposures along the northern semi-major axis), the major-axis
kinematics of NGC147 now extend out to $\sim 150''$ (0.5~kpc or
1.2~$R_{\rm e}$) along the northern semi-major axis and out to $\sim
100''$ (0.3~kpc or 0.8~$R_{\rm e}$) along the southern semi-major
axis.

The details of the modeling method are given in Section \ref{3int}. We
present the results for each of the galaxies in Sect.
\ref{result}. Our conclusions are summarized in Sect. \ref{conc}.

\section{Three-integral dynamical models} \label{3int}

The internal dynamics of a steady-state oblate stellar system are
described by a gravitational potential $\psi(\varpi,z)$, that
determines the stellar orbits, and the distribution function (DF)
$F(\vec{r},\vec{v})\,d\vec{r}\,d\vec{v}$, which gives the number
density of stars in phase space ($(\varpi,z)$ are cylindrical
coordinates in each meridional plane). Loosely speaking, the DF
distributes the stars over all possible orbits in a given
potential. In the following, we will work in prolate elliptical
coordinates $(\lambda,\nu$) \cite{dz88}, defined by
\begin{equation}
\varpi^2 = \frac{(\lambda+\alpha)(\nu+\alpha)}{\alpha-\gamma}, \, 
z^2 = \frac{(\lambda+\gamma)(\nu+\gamma)}{\gamma-\alpha},
\end{equation}
with $-\gamma \le \nu \le -\alpha \le \lambda$. Surfaces of constant
$\lambda$ are confocal prolate spheroids; surfaces defined by constant
$\nu$ are confocal two-sheeted hyperboloids, both with foci at $z =
\pm \Delta = \pm \sqrt{\gamma - \alpha}$. We will approximate the
gravitational potential by a St\"ackel potential, of the form
\begin{equation}
\psi(\varpi,z) = \frac{(\lambda+\gamma) G(\lambda) - (\nu+\gamma)
G(\nu)}{\lambda-\nu},
\end{equation}
with $G(\lambda)$ the potential in the equatorial plane. Such
potentials allow the existence of three integrals of motion: the
binding energy $E$, the $z$-component of the angular moment $L_z$,
which we use in the more convenient form $I_2 = L_z^2/2$, and
\begin{equation}
I_3 = \frac{1}{2} (L^2 - L_z^2) + (\gamma - \alpha)\left( \frac{1}{2}
v_z^2 - z^2 \frac{G(\lambda) - G(\nu)}{\lambda-\nu} \right),
\end{equation}
an integral which is a generalisation of angular momentum conservation
in spherical systems. Roughly, for a given $E$ and $I_2$, $I_3$
determines how high above the equatorial plane an orbit will
come. Models that make use of the third integral are called
three-integral or 3I models, versus two-integral or 2I models that
go without it.

A detailed account of the method we employed to construct the
spheroidal coordinate system and St\"ackel potential that give the
best fit to a given axisymmetric potential can be found in Dejonghe \&
de Zeeuw \shortcite{dz88}, Dejonghe et al. \shortcite{de96}, and in De
Bruyne et al. \shortcite{deb01}. In brief, we deproject the observed
surface brightness distribution, derived from a 2MASS $J$-band image,
assuming the galaxy to be axisymmetric. Since NGC147 and NGC205 have a
fairly flattened apparent shape, with an apparent axial ratio $q_{\rm
app}=0.5$, it is quite unlikely that they are viewed far from
edge-on. The intrinsic flattening distribution of dEs peaks around an
intrinsic axial ratio $q_{\rm intr}=0.6$ while galaxies with an
intrinsic axial ratio smaller than $q_{\rm intr}=0.4$ are virtually
absent \cite{bp95}. Assuming both galaxies to have an oblate,
axissymetric light distribution that is intrinsically rounder than a
$q_{\rm intr}=0.4$ shape, the inclination is expected to be larger
than $i \sim 70^{\circ}$. Therefore, we can assume for simplicity that
NGC147 and NGC205 are viewed edge-on without significantly affecting
any of the results. NGC185, on the other hand, has a much rounder
apparent shape, with an apparent axial ratio $q_{\rm app}=0.8$. If we
assume that the galaxy has an intrinsic shape that is rounder than
$q_{\rm intr}=0.4$, the inclination is expected to be larger than $i
\sim 40^{\circ}$. In order to assess the influence of the intrinsic
shape and the viewing angle of NGC185 on our results, we generated
models with $i=90^{\circ}$, $i=60^{\circ}$, $i=55^{\circ}$,
$i=50^{\circ}$, and $i=45^{\circ}$. However, with kinematics along the
major axis only and lacking higher-order moments of the line-of-sight
velocity distribution, the kinematical data do not allow to constrain
the inclination with any degree of precision.

The total mass density, including dark matter, is parameterized as the
spatial luminous mass density multiplied by a spatially varying
mass-to-light ratio
\begin{equation}
\frac{M}{L}(\varpi,z) = A \left(1 + B \sqrt{\varpi^2 + (z/q)^2}\right), \label{pot}
\end{equation}
with the parameters $A$ and $B$ to be estimated from the data and $q$
the axis ratio of the luminosity density distribution. Attempts to
find intermediate-mass black holes in dEs, using either ground-based
kinematics \cite{g02} or high spatial resolution kinematical
information obtained with HST \cite{v05}, have sofar failed. For this
reason, and since we are mostly interested in the total dark-matter
content of these galaxies, we did not include a central black hole in
our models. The gravitational potential is obtained by decomposing the
total mass density in spherical harmonics. Finally, we determine the
St\"ackel potential, i.e. the focal length $\Delta$ and the function
$G$, that best fits this gravitational potential. The St\"ackel form
of the potential is used only to calculate the third integral $I_3$;
elsewhere the potential derived from spherical harmonics is used. The
potential is normalised such that $\psi(r) = 1/r$ at large radius $r$,
with $r$ expressed in kiloparsecs.

For a given potential, we wish to find the DF that best reproduces the
kinematical information. As a first step, the DF is written as a
weighted sum of basis functions, called ``components''. Here, we will
use components of the form
\begin{eqnarray}
F^{i,\epsilon_i}(E,I_2,I_3) &=& (E - E_{0,i})^{\sigma_i} (I_2 -
I_{0,i})^{\tau_i} I_3^{\rho_i}, \nonumber \\ && \hspace{3.93em}\,{\rm
if}\,\,E>E_{0,i},\,\epsilon_i I_2 > I_{0,i} \ge 0 \nonumber \\ &=& 0,
\hspace{3em}\,{\rm if}\,E \le E_{0,i}\,\,{\rm or}\,\,\epsilon_i I_2 \le I_{0,i}
\end{eqnarray}
with $\sigma_i$, $\tau_i$, and $\rho_i$ integer numbers. $E_{0,i}$ is
a lower bound on the binding energy, which defines the outer boundary
within which the component is non-zero. The parameter $\epsilon_i$
indicates whether stars are placed only on orbits with positive angular
momentum ($\epsilon_i=+1$) or only on orbits with negative angular
momentum ($\epsilon_i=-1$) or on both ($\epsilon_i=0$). We refer the
reader to Appendix \ref{app} for a discussion of the properties of
these components. The DF then takes the form $F(E,I_2,I_3) = \sum_{i}
c_{i} \, F^{i,\epsilon_i}(E,I_2,I_3)$. The coefficients $c_{i}$ are
determined by minimizing the quantity
\begin{equation}
\chi^2 = \sum_l \left( \frac{ {\rm obs}_l - \sum_{i} c_{i} \,{\rm
obs}_l^{i} }{\sigma_l} \right)^2, \label{chi2}
\end{equation}
with ${\rm obs}_l$ an observed data point, $\sigma_l$ the 1-$\sigma$
errorbar on that data point, and ${\rm obs}_l^{i}$ the corresponding
value calculated from the basis function $F^{i,\epsilon_i}$, subject
to the constraint that the DF be positive everywhere in phase
space. Note that the coefficients $c_i$ can be negative, as long as
the weighted sum of basis functions is positive.

The projected velocity moments, $\mu_n(\Omega)$ can be obtained by
integrating suitably weighted combinations of the spatial velocity
moments over a line of sight in a direction $\Omega$:
\begin{eqnarray}
\mu_n(\Omega) &=& \int_\Omega v_p^n F(E,I_2,I_3)\,d\vec{v}\,d\xi \nonumber \\
&=& \sum_i c_i \int_\Omega  v_p^n F^{i,\epsilon_i}(E,I_2,I_3)\,d\vec{v}\,d\xi
\end{eqnarray}
with $v_p$ the line-of-sight velocity and $\xi$ a coordinate along the
line of sight. From the observed quantities, the projected luminosity
density $\rho_p = 10^{(25.26 - \mu_J)/2.5}\,L_{J,\odot}{\rm pc}^{-2}$
with $\mu_J$ the observed surface brightness expressed in
magnitude~arcsec$^{-2}$, the mean projected velocity $v_p$, and the
projected velocity dispersion $\sigma_p$, one can easily construct
observed values for the lowest-order velocity moments $\mu_0 =
\rho_p$, $\mu_1 = \rho_p\,v_p$, and $\mu_2 = \rho_p (v_p^2 +
\sigma_p^2)$ along each observed line of sight and insert these values
into eqn. (\ref{chi2}), which is minimized using a Quadratic
Programming algorithm \cite{d89}. The minimisation algorithm has a
library at its disposal that contains hundreds of components to choose
from. Since the algorithm itself determines which and how many
components are used, this is a virtually non-parametric method of
retrieving the distribution function. Typically, the $\chi^2$ does not
further improve after co-adding a few tens of components and the
procedure can be stopped. This finally yields the coefficients $c_i$
and consequently the distribution function that best reproduces the
data for a given gravitational potential. If the basis functions
$F^{i,\epsilon_i}$ are smooth and well-behaved functions of the
integrals of motion, any combination of a finite number of basis
functions will by construction also be smooth and well-behaved in
phase space.

It is well known that in a spherical geometry even complete
kinematical knowledge does not suffice to uniquely constrain both the
gravitational potential and the distribution function. On the contrary
: for each potential, there is a unique distribution function that
reproduces the data \cite{dm92}. That distribution function, however,
need not be positive everywhere in phase-space. The positivity
constraint we are imposing on the distribution function therefore
helps to constrain the gravitational potential. The degeneracy in the
general axisymmetric case, however, can be expected to be a lot
smaller than in the spherical case. For 2I models, the even part of
the distribution function can be derived directly from the augmented
stellar density $\tilde{\rho}(\varpi,\psi)$ which, in the absence of
dark matter, can in principle be constructed directly by deprojecting
the observed surface brightness distribution. This is clearly not
possible in the spherical case. Therefore, even if the kinematical
data are restricted to the three lowest order moments of the
distribution function along the major (and sometimes minor) axis, we
can expect to be able to put meaningful constraints on the dark matter
content and mass, a point already made in Dejonghe et
al. \shortcite{de96}. 3I models have a larger freedom in distributing
stars over all possible orbits than 2I models, widening somewhat the
range of possible solutions that is consistent with the data.

\begin{figure}
\vspace*{7cm}
\special{hscale=55 vscale=55 hsize=500 vsize=500
hoffset=-20 voffset=-110 angle=0 psfile="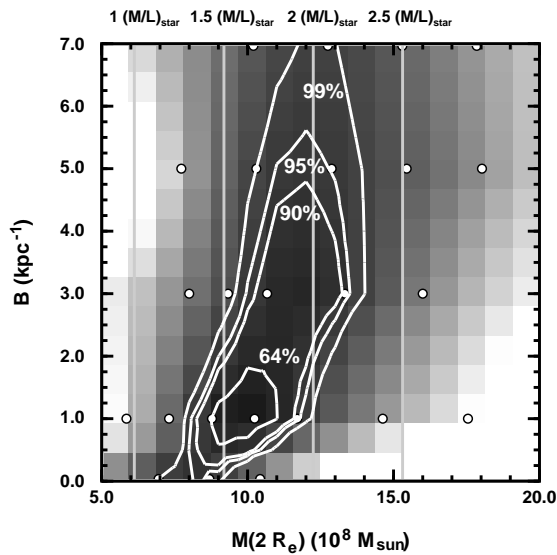"}
\caption{Contours of constant $\chi^2$ for the models fitted to the
NGC205 data as a function of the total mass within a 2~$R_{\rm e}$
radius, $M(2\,R_{\rm e})$, and the scale parameter $B$ of equation
(\ref{pot}). The white dots mark the parameter values of the models we
constructed. The contours indicate the 64\%, 90\%, 95\%, and 99\%
confidence levels. At the 95\% confidence level, the mass within a
2~$R_{\rm e}$ radius is $M(2\,R_{\rm e}) = 10.2^{+3.3}_{-2.2} \times
10^8 \,M_\odot$. For the scale parameter, we find $B
=1.1^{+4.5}_{-1.1}$~kpc$^{-1}$, at the same confidence level. The best
fitting model evidently requires a spatially varying mass-to-light
ratio to reproduce the data. Models with a constant mass-to-light
ratio, i.e. with $B=0$~kpc$^{-1}$ are barely acceptable at the 95\%
confidence level. The vertical lines indicate the locus of models for
which the mass-to-light ratio at 2~$R_{\rm e}$ equals 1, 1.5, 2, and
2.5 times the estimated stellar mass-to-light ratio ($(M/L)_{\rm J} =
1.32$, see section \ref{dynmass}). All acceptable models, even the
ones with a zero $B$, have mass-to-light ratios well above that of a
purely stellar system.
\label{B_M_chi2}}
\end{figure}

\begin{figure*}
\vspace*{10cm}
\special{hscale=95 vscale=95 hsize=500 vsize=290
hoffset=-20 voffset=-215 angle=0 psfile="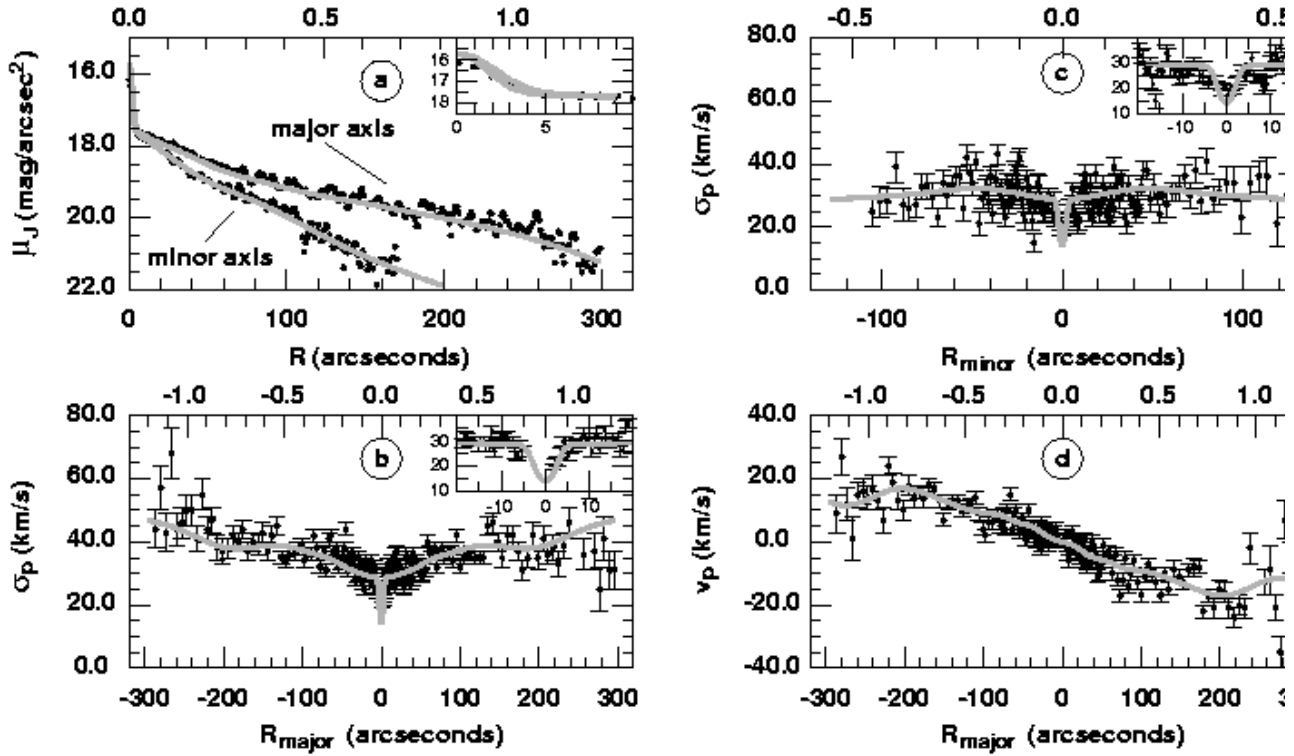"}
\caption{Fit to the kinematics of NGC205. Panel {\bf a}:~the J-band
surface brightness, $\mu_J$, along major and minor axis; panel {\bf
b}:~the major-axis velocity dispersion, $\sigma_p$; panel {\bf c}:~the
minor-axis velocity dispersion; panel {\bf d}:~the mean streaming
velocity along the major axis, $v_p$. The black dots are the data
points, the grey curves correspond to the best fitting dynamical
model. The top axis of each panel is labeled with radius expressed
in kiloparsecs. The model was fitted to the surface brightness on a
logarithmically spaced grid covering the whole face of the galaxy but
for clarity we only plot the major and minor axis profiles in this
figure. The model accurately reproduces the central surface brightness
peak (see insets in panel {\bf a}), and the corresponding central drop
in the velocity dispersion (see insets in panels {\bf b} and {\bf
c}). The major-axis velocity dispersion appears to keep rising very
slowly out to the last data point, at 5{\arcmin}. \label{kin205}}
\end{figure*}

\begin{figure*}
\vspace*{9cm}
\special{hscale=65 vscale=65 hsize=700 vsize=250
hoffset=-45 voffset=-20 angle=0 psfile="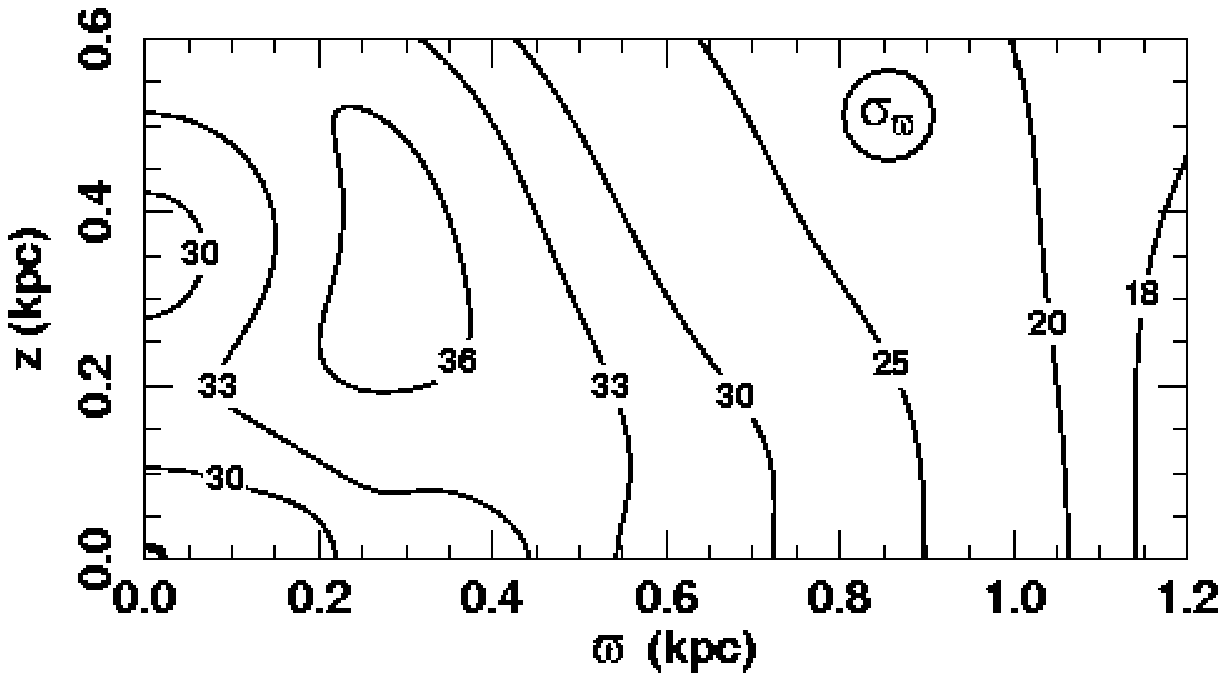"}
\special{hscale=65 vscale=65 hsize=700 vsize=250
hoffset=225 voffset=-20 angle=0 psfile="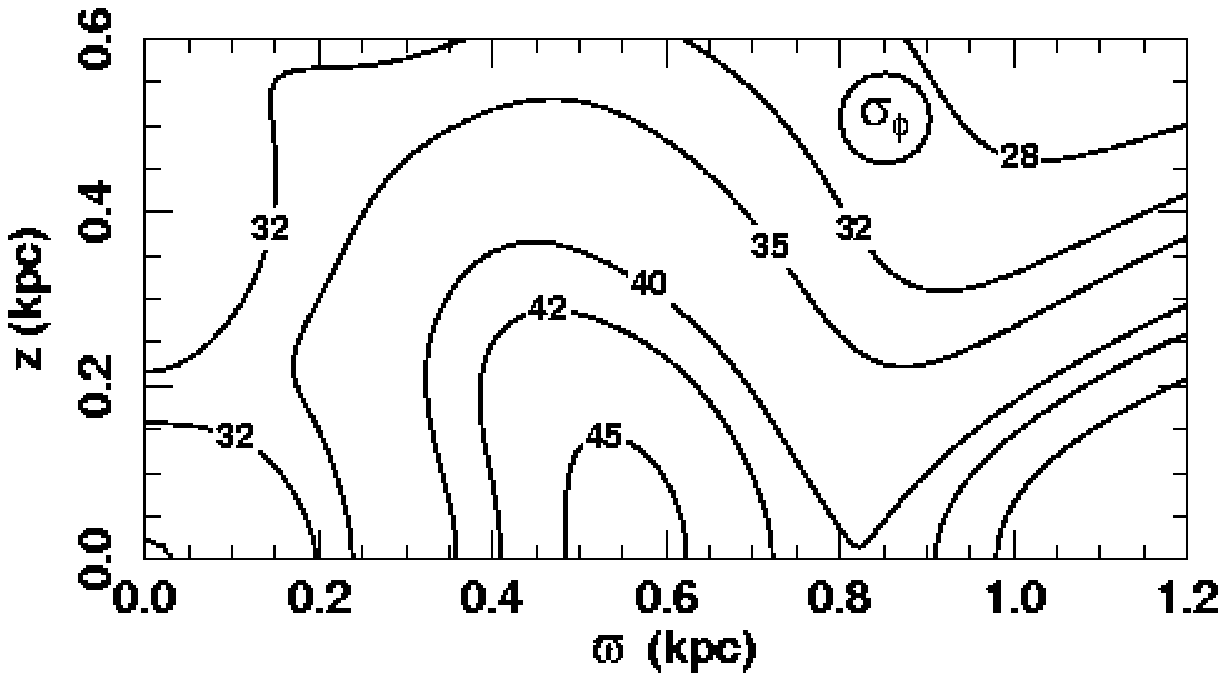"}
\special{hscale=65 vscale=65 hsize=500 vsize=130
hoffset=-45 voffset=-150 angle=0 psfile="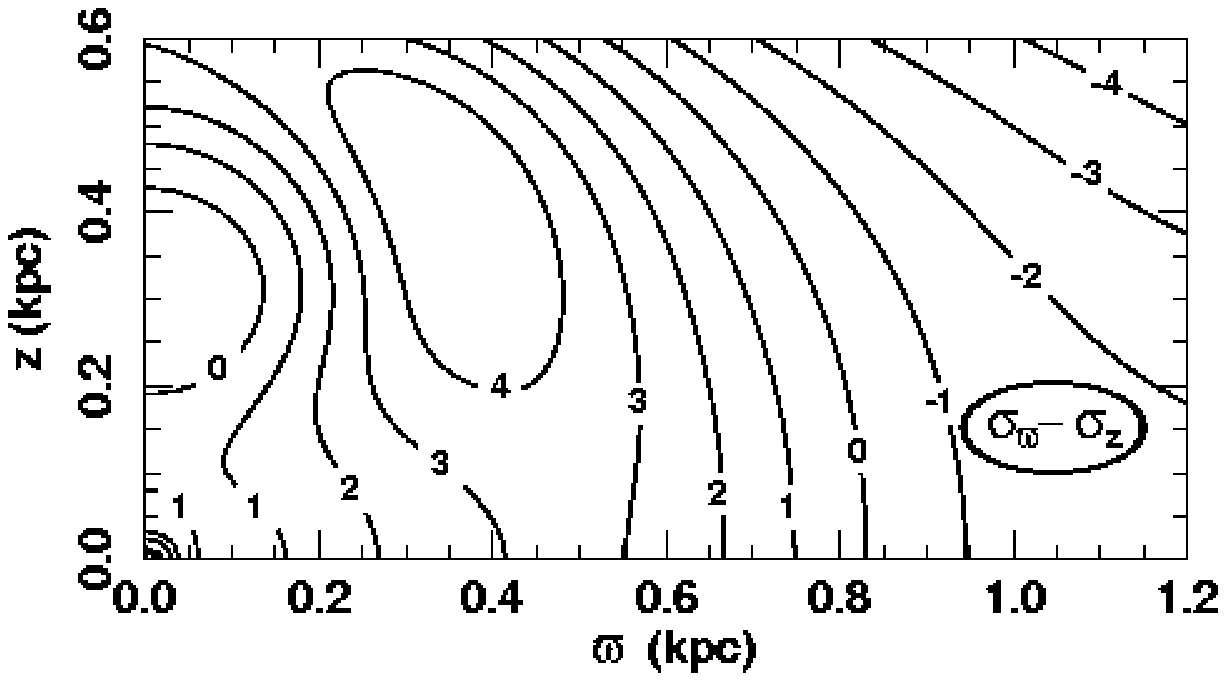"}
\special{hscale=65 vscale=65 hsize=500 vsize=130
hoffset=225 voffset=-150 angle=0 psfile="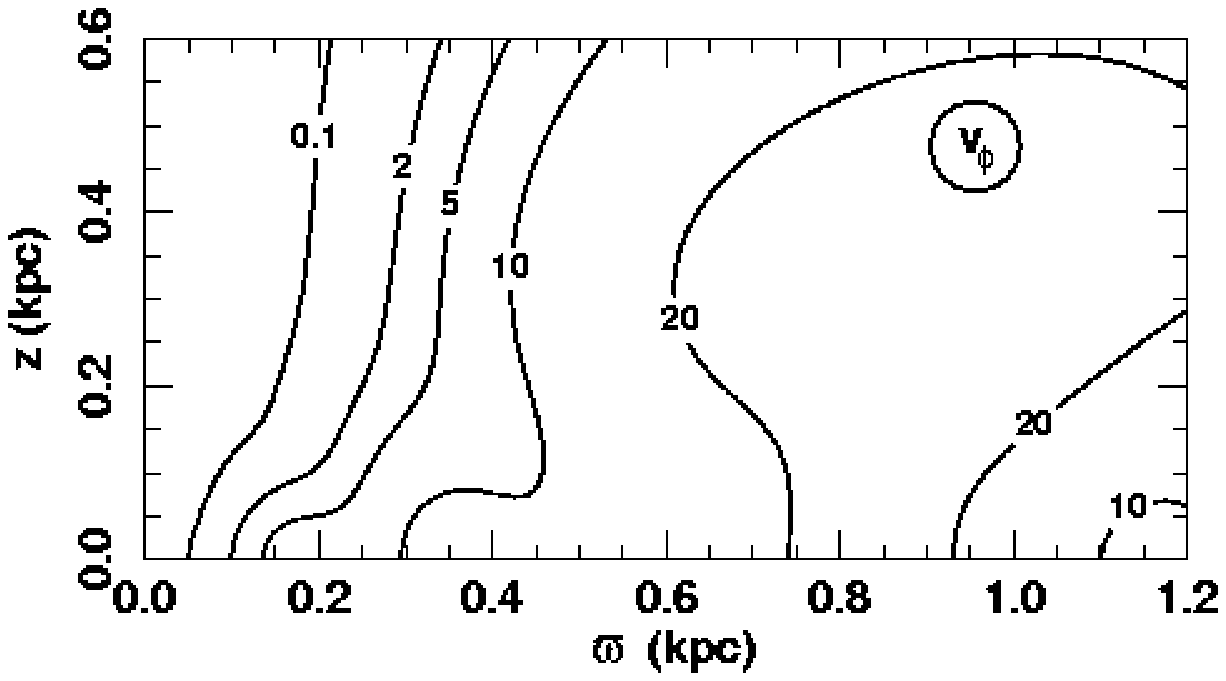"}
\caption{Contours for the radial velocity dispersion, $\sigma_\varpi$,
the tangential velocity dispersion, $\sigma_\phi$, the difference
between the radial and the vertical velocity dispersion $\sigma_\varpi
- \sigma_z$, and the mean tangential velocity, $v_\phi$, in the
meridional plane of the best fit dynamical model for NGC205. All
contours are labeled in km~s$^{-1}$. \label{spatkin205}}
\end{figure*}

We fitted models to the data of each galaxy with various values for
the parameters $A$ and $B$ in eqn. (\ref{pot}). The model with the
lowest $\chi^2$-value, which most closely reproduces the data, is
retained as the best model. The uncertainties on the dynamical
quantities presented in this paper are derived from the range of
allowed models (all models within the 95\% confidence level are deemed
acceptable). 

\section{Results} \label{result}

\subsection{NGC205}

\subsubsection{New and previous kinematical work}

The major and minor axis J-band photometry and kinematics of NGC205 as
measured by Simien \& Prugniel \shortcite{sp02} are presented in
Fig. \ref{kin205}. These data were derived from spectra with a
25~km~s$^{-1}$ velocity resolution, using a Fourier-Fitting
technique. Detailed simulations showed that template mismatch has
little effect on the kinematical parameters \cite{ko06}. There is a
slight degeneracy between the velocity dispersion and the metallicity:
a template of too high a metallicity will give too large a
dispersion. The template star we used has a metallicity [Fe/H]$=-0.5$
\cite{ps01}. Our simulations show that for a galaxy with [Fe/H]$=-1$
and a template with [Fe/H]$=-0.5$, the dispersion can be overestimated
by $\sim 12$\% at most. However, the fitting procedure includes an
additive continuum which decreases the mismatch by an order of
magnitude. Age mismatch, as is likely to be the case in the peculiar
nucleus of NGC205, is not expected to bias the kinematics.

NGC205 has a very steep central density cusp or nucleus
(see panel {\bf a} of Fig. \ref{kin205}). Within the inner 3
arcseconds, the J-band surface brightness becomes almost 2
mag~arcsec$^{-2}$ brighter. This sudden density jump is reflected in a
corresponding drop of the velocity dispersion by about 10~km~s$^{-1}$
towards the center; from $\sigma_p = 30 \pm 2$~km~s$^{-1}$ at a radius
of 10$''$ down to a central velocity dispersion of $\sigma_p = 20 \pm
1$~km~s$^{-1}$ (see panels {\bf b} and {\bf c} of
Fig. \ref{kin205}). The nucleus of NGC205 therefore appears to be a
round and very dense but dynamically cold substructure. Outside the
nucleus, the velocity dispersion was found to rise outwardly from
30$\pm 2$~km~s$^{-1}$ near the center up to about 45$\pm
7$~km~s$^{-1}$ at 300{\arcsec}. The rotation velocity rises to a
maximum of 20$\pm 5$~km~s$^{-1}$ at 200{\arcsec} and declines again
beyond that radius. This was the first clear detection of rotation in
this galaxy. There is evidence that the mean velocity reverses sign at
a radius of about 300{\arcsec}. This feature agrees with other
observations \cite{h73,g04} that suggest that the tidal forces of M31
only appreciably affect the outer regions of NGC205, beyond
5{\arcmin}. Moreover, the absence of minor-axis rotation \cite{sp02}
and the presence of a very dense nucleus, which is likely to scatter
stars off box-orbits, which are the backbone of any slowly rotating
triaxial mass-distribution, into loop-orbits \cite{ec94,mq98} both
argue against NGC205 having a strongly triaxial mass distribution,
justifying our assumption of an oblate geometry for the construction
of dynamical models.

\begin{figure*}
\vspace*{6cm} \special{hscale=48 vscale=48 hsize=500
vsize=180 hoffset=297 voffset=-104 angle=0 psfile="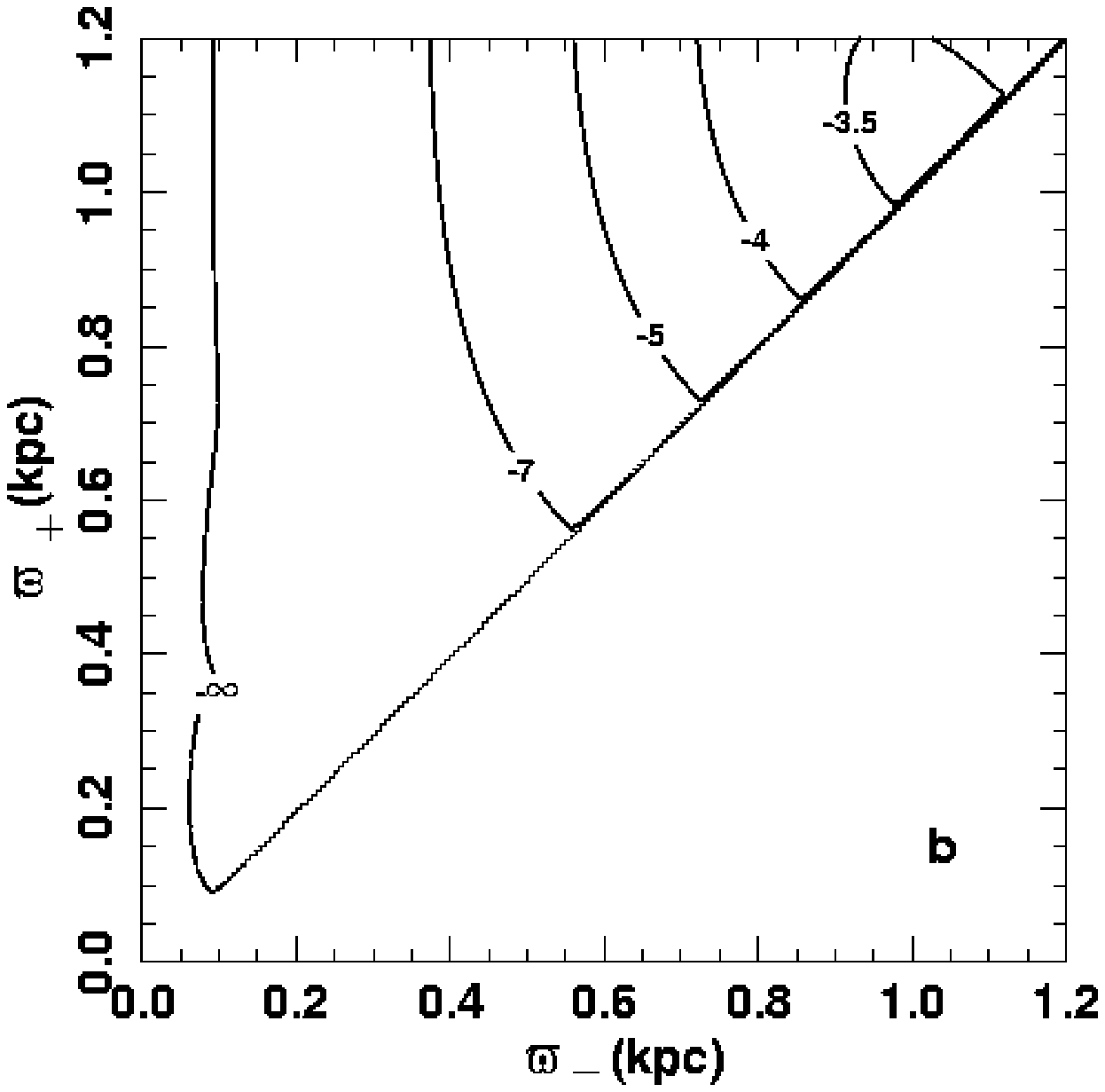"}
\special{hscale=57.6 vscale=57.6 hsize=500
vsize=180 hoffset=-34 voffset=-130 angle=0 psfile="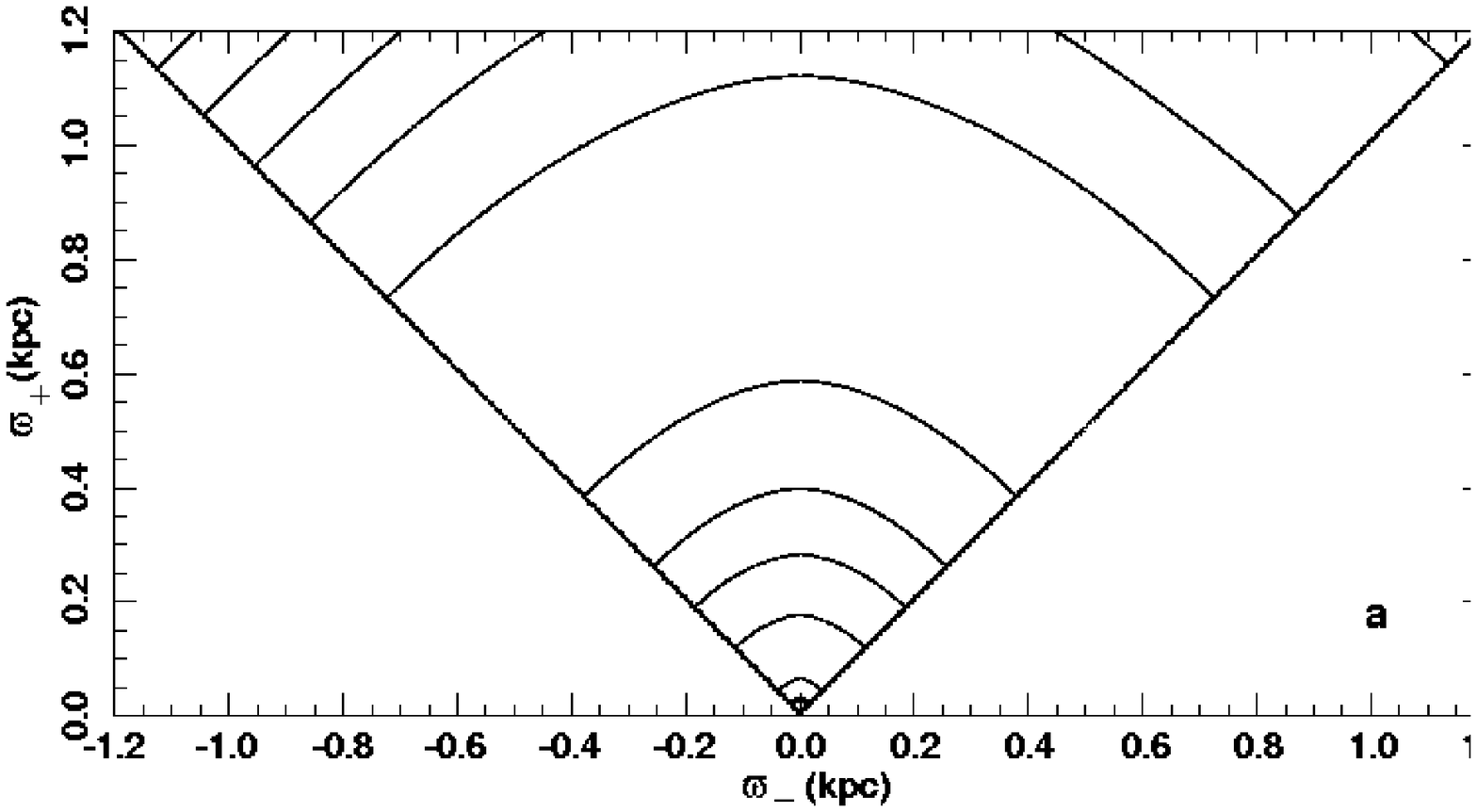"}
\caption{Panel {\bf (a)}:~the distribution function of the best fit
dynamical model for NGC205, plotted in turning-point space for loop
orbits in the equatorial plane, characterized by $I_3=0$. The
pericenter distance is denoted by $\varpi_-$; the apocenter distance
by $\varpi_+$. Radial orbits populate the $\varpi_-=0$ axis while
circular orbits lie along the diagonal. $\varpi_-$ has the same sign
as the angular momentum $L_z$. The contour labels are expressed in
units of $\log( L_{\odot,\rm J} \,{\rm pc}^{-3}\, ({\rm
50\,km~s^{-1}})^{-3})$, from $-4.5$ to $1.0$ with a 0.5 interval. The
orbital distribution within the equatorial plane is clearly very
nearly isotropic, with only a very slight excess of stars with
positive angular momentum $L_z$, causing the rotation of NGC205. Panel
{\bf b}:~to have a more detailed look at this slight excess of stars
on positive angular momentum orbits, we plotted $F(E,L_z > 0,0) -
F(E,L_z<0,0)$. The contour labels are expressed in units of $\log(
L_{\odot,\rm J} \,{\rm pc}^{-3}\, ({\rm 50\,km~s^{-1}})^{-3})$. At
small radii, there is almost exact counter-rotation, with stars on
positive and negative angular momentum orbits almost cancelling any
rotation. At larger radii, the number of stars on positive angular
momentum orbits starts to dominate and at a radius of $\sim 1$~kpc,
there are about twice as many stars on right-handed near-circular
orbits as there are on left-handed ones. \label{DFTP_205}}
\end{figure*}

Previous work on the kinematics of NGC205 has yielded contradictory
results. Held et al. \shortcite{he90} present major axis velocity and
velocity dispersion profiles out to about 1{\arcmin}, measured from
spectra with a 83~km~s$^{-1}$ per pixel spectral resolution and using
a G5 star spectrum as template. The mean velocity dispersion is about
60-70~km~s$^{-1}$, with only a very slight decline towards the center,
reaching about 50~km~s$^{-1}$ in the very center. No significant
rotation was detected. These authors deduce a mass-to-light ratio of
$(M/L)_{\rm B} \sim 7\,M_\odot/L_{\odot,\rm B}$. Bender et
al. \shortcite{be91} on the other hand found the velocity dispersion
to rise almost linearly with radius, from $\sim 30$~km~s$^{-1}$ near
the center to $\sim 70$~km~s$^{-1}$ at one half-light radius. At the
very center, the velocity dispersion was found to drop to
20~km~s$^{-1}$. Again, rotation turned out to be negligible. The
kinematics were derived at a 30~km~s$^{-1}$ spectral resolution using
a single G8{\sc iii} stellar template spectrum. These authors estimate
the B-band mass-to-light ratio at $(M/L)_{\rm B} = 8 \pm 2
\,M_\odot/L_{\odot,\rm B}$. In Held et al. \shortcite{he92}, the mean
value of the major-axis velocity dispersion of NGC205, $\langle \sigma
\rangle = 42$~km/s, is used to arrive at $(M/L)_{\rm B} \sim
3.5\,M_\odot/L_{\odot,\rm B}$. Carter \& Sadler \shortcite{cs90} on
the other hand retrieved an almost flat velocity dispersion out to
1{\arcmin}, constant at about 45~km~s$^{-1}$. Very near the center,
the velocity dispersion was observed to drop to $\sim
15$~km~s$^{-1}$. These kinematics were measured using a composite
stellar template template from a galaxy spectrum with a very high
spectral resolution (5~km~s$^{-1}$). These authors estimate the
mass-to-light ratio at $(M/L)_{\rm B} \sim 9.4 \,M_\odot/L_{\odot,\rm
B}$ with an uncertainty of a factor of 2.

\subsubsection{The dynamical mass and mass-to-light ratio} \label{dynmass}

From the range of models that is compatible with the data at the 95\%
confidence level, we estimate the mass within a 2~$R_{\rm e}$ radius
sphere at $M(2\,R_{\rm e}) = 10.2^{+3.3}_{-2.2} \times 10^8
\,M_\odot$, corresponding to a J-band mass-to-light ratio within the
inner 2~$R_{\rm e}$ of $(M/L)_{\rm J}
=2.2^{+0.7}_{-0.4}\,M_\odot/L_{\odot,\rm J}$ or a B-band mass-to-light
ratio of $(M/L)_{\rm B} = 4.5^{+1.5}_{-1.0}\,M_\odot/L_{\odot,\rm B}$,
using a B$-$J$=2.397$ color, corrected for Galactic reddening
\cite{rc3,ja03}. The mean metallicity of NGC205 is quite well
determined at [Fe/H$]\approx -0.9$, using various techniques
\cite{mc05,bm05,ri98}. In the central regions of our dynamical models,
mass follows light. If the central mass-to-light ratio of the best fit
model, $(M/L)_{\rm J} = 1.32\,M_\odot/L_{\odot,\rm J}$, is indicative
of that of the stellar population and if we adopt a mean metallicity
[Fe/H$]\approx -0.9$, we estimate the mean age of the stellar
population at $\sim 9$~Gyr \cite{wo94}, which is not unreasonable. The
model with the lowest mass that is still consistent with the data at
the 95\% confidence level has a constant mass-to-light ratio
$(M/L)_{\rm J} = 1.8\,M_\odot/L_{\odot,\rm J}$, which is significantly
higher than that estimated for the stellar population above. Hence, we
can state that NGC205 is {\em not} a pure stellar system, completely
devoid of dark matter. Within a $2\,R_{\rm e}$ sphere, NGC205 consists
of about 60\% luminous matter, by mass, and of 40\% dark matter.

\subsubsection{The internal dynamics}

\begin{figure}
\vspace*{14.5cm}
\special{hscale=90 vscale=90 hsize=500 vsize=400
hoffset=-30 voffset=-70 angle=0 psfile="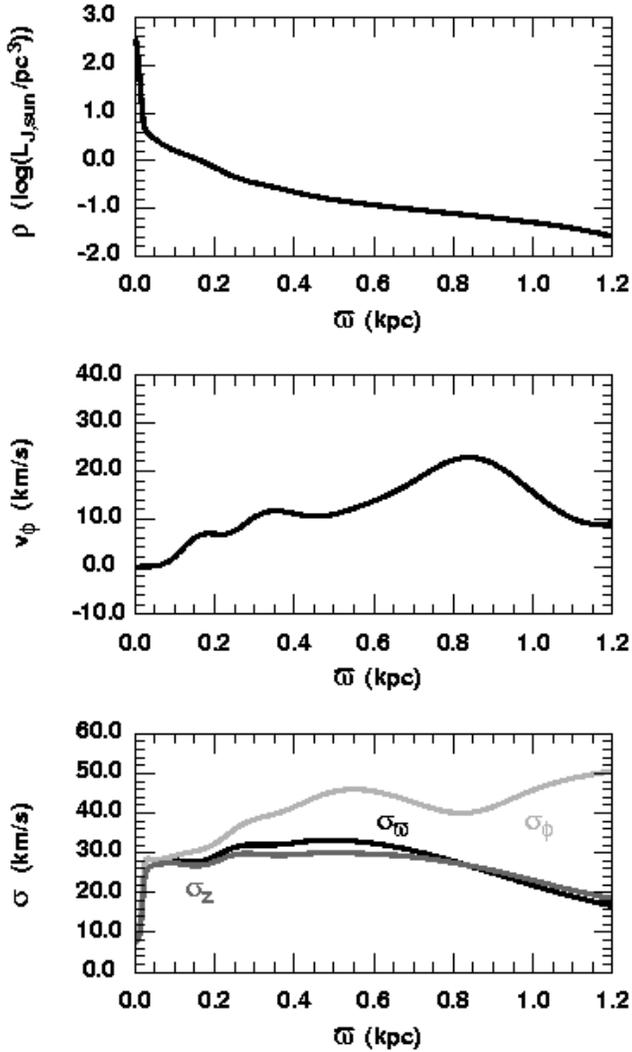"}
\caption{The internal dynamics of the best fit model for NGC205,
plotted as a function of radius within the equatorial plane. Top
panel:~the logarithm of the luminosity density of the stars, $\rho$;
middle panel:~the rotation velocity, $v_\phi$; bottom panel:~the
radial, tangential, and vertical components of the velocity dispersion
tensor. \label{spc205}
}
\end{figure}
The best fitting 3I model for NGC205 very accurately reproduces the
photometric and kinematical data, as can be judged from
Fig. \ref{kin205}, where the kinematics of the best fitting model are
overplotted onto the data. Since the dynamics of this galaxy are
constrained by kinematical information along both the major and minor
axes, it is instructive to more closely examine some internal
dynamical quantities. In Fig. \ref{spatkin205}, the spatial kinematics
are presented in the meridional plane:~the radial velocity dispersion
$\sigma_\varpi$, the tangential velocity dispersion $\sigma_\phi$, the
difference between the radial and vertical velocity dispersions
$\sigma_\varpi - \sigma_z$ (in the case of a 2I model:~$\sigma_\varpi
- \sigma_z = 0$), and the mean tangential velocity $v_\phi$.

Geometrically, it is clear that the minor-axis velocity dispersion
puts strong constraints on the behaviour of the radial velocity
dispersion along the symmetry axis while the major-axis kinematics
constrain both the radial and the tangential velocity dispersion
within the equatorial plane. The vertical velocity dispersion is only
very weakly constrained by the kinematical data; it will mostly be
determined by the photometry, i.e. by the flattening of the stellar
distribution. As is clear from Figs. \ref{kin205}, \ref{spatkin205},
and \ref{spc205}, the behaviour of the major-axis projected velocity is
reflected in that of the tangential component $\sigma_\phi$ of the
velocity dispersion tensor:~the maximum around 0.5~kpc (125{\arcsec})
and the ourward rise beyond 0.8~kpc (200\arcsec) can be clearly
identified. The run of the radial velocity dispersion along the
$z$-axis closely traces the minor-axis projected velocity dispersion,
as expected; cf. the maximum around 0.2~kpc (50\arcsec). In a 3I
model, the radial and vertical velocity dispersions are
decoupled. Still, the radial versus vertical anistropy is very
small~:~the vertical velocity dispersion is only marginally smaller
than the radial dispersion within the inner 0.8~kpc (200\arcsec). If
we use the anisotropy parameter $\beta_{\varpi,z} =
1-(\sigma_z/\sigma_\varpi)^2$ to quantify this anisotropy, then we
find that $-0.6 < \beta_{\varpi,z} < 0.3$ everywhere in the meridional
plane. The radial versus tangential anisotropy is somewhat larger. If we
introduce the quantity $\beta_{\varpi,\phi} =
1-(\sigma_\phi/\sigma_\varpi)^2$, then we find that $-1 <
\beta_{\varpi,\phi} < 0.1$ within the region that is well constrained
by the kinematical data. Beyond a radius of 0.8~kpc, the tangential
anisotropy increases to keep up with the observed rise of the
projected velocity dispersion along the major axis.

The mean tangential velocity $v_\phi$ rises only very slowly as a
function of radius, as does the projected mean velocity $v_p$ along
the major axis. The velocity reaches a maximum at 0.8~kpc (200\arcsec)
and declines beyond that radius (see Fig. \ref{spc205}). Extrapolating
somewhat beyond the extent of our data, the velocity is expected to
reverse sign at about 300{\arcsec}, in agreement with Geha
\shortcite{g04}. From Fig. \ref{DFTP_205}, it is clear that there is
only a slight excess of stars with positive angular momentum $L_z$
causing the observed rotation. Only beyond $\sim 0.7$~kpc does the
number of stars with $L_z > 0$ clearly dominate that of $L_z < 0$
stars. At a radius of $\sim 1$~kpc, there are about twice as many
stars on right-handed near-circular orbits as there are on left-handed
ones.

NGC205 has a very distinct nucleus, which shows up in the dynamical
model as a spherically symmetric, dynamically very cold substructure
(see Figs. \ref{spatkin205} and \ref{spc205}). By singling out those
basis functions for the distribution function that make up the
nucleus, i.e. the basis functions with a very small outer boundary, we
can study the nucleus in more detail. The nucleus can be modeled as an
isotropic star cluster with a one-dimensional velocity dispersion
$\sigma \approx 7$~km~s$^{-1}$ and a truncation radius of about
50~pc. By integrating the spherically symmetric mass density of the
nucleus, we estimate its total mass at $1.4 \times
10^6\,M_\odot$. This is in excellent agreement with a previous
estimate based on a King-model fit to the surface brightness profile
of the nucleus and its projected velocity dispersion
\cite{cs90,jo96}. It should be noted, however, that the mass-to-light
ratio of the nucleus is probably different from that of the main body
of NGC205 \cite{v05}. The functional form for the mass-to-light ratio
adopted by us, eq. (\ref{pot}), cannot reproduce a central
$M/L$-variation. However, this is not our intention:~we are interested
in quantifying the amount of dark matter at large radii. Since the
nucleus contributes only of the order of 0.1~\% of the total mass of
the galaxy, our total mass estimates are robust.

Thus, the nucleus of NGC205 is structurally and dynamically not unlike
a massive globular cluster. The orbital decay time scale for a
globular cluster in NGC205 by dynamical friction is of the order
$10^8-10^9$ years, significantly less than a Hubble time
\cite{lo01,jo96,tos75}, which makes it plausible that one or more star
clusters have spiraled inwards and settled at the bottom of the
gravitational well. Though detailed N-body simulations of this process
are still lacking, the observed properties of the NGC205 nucleus are
not inconsistent with this scenario.

\begin{figure}
\vspace*{8cm}
\special{hscale=90 vscale=90 hsize=500 vsize=240
hoffset=-60 voffset=-160 angle=0 psfile="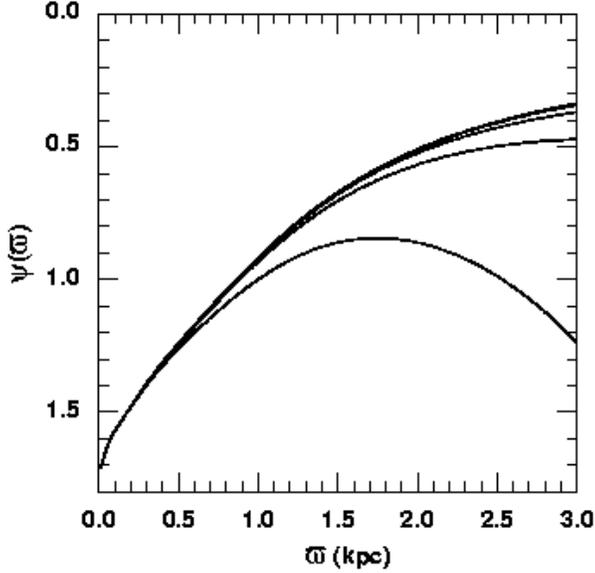"}
\caption{The gravitational potential well $\psi$ of NGC205 in the
equatorial plane as a function of radius $\varpi$. The top curve is
the gravitational potential of the best fit dynamical model for NGC205
in the abscence of M31. From top to bottom, the other curves indicate
the ``effective'' potential in the presence of the axisymmetrized
tidal field of M31 (eq. (\ref{poteq})) for relative distances of 40
(which essentially overlaps with the unperturbed potential of NGC205),
20, 10, and 5~kpc between NGC205 and M31.
\label{pot205}}
\end{figure}

\subsubsection{The tidal influence of M31}

The TRGB distance estimates from McConnachie \shortcite{mc05} place
NGC205 $48 \pm 33$~kpc behind M31. The 1$\sigma$ error on the TRGB
estimates for the distances of NGC205 and M31 are large enough that it
is quite plausible that NGC205 and M31 are at the same distance from
the Milky Way and hence at a relative distance of a mere 10~kpc.

In order to get a rough estimate of how the presence of M31 affects
our results, we approximate the gravitational potential of M31 as that
of a spherical mass distribution at a distance $R$ from NGC205. For
computational ease, we will further assume that the center of mass of
M31 lies within the equatorial plane of NGC205 and that distances
within the stellar body of NGC205 are much smaller than $R$. We then
axisymmetrize this potential by averaging it over the azimuthal angle
$\phi$. We finally obtain the expression
\begin{eqnarray}
\Phi_{\rm tidal, M31}(\varpi,z) &=& -\frac{G M(R)}{\varpi}\left(
\frac{1}{4} \left(\frac{\varpi}{R}\right)^2 -
\frac{1}{2}\left(\frac{z}{R}\right)^2 \right. \nonumber \\ &&
\hspace{-8em} \left.  + \frac{9}{64} \left(\frac{\varpi}{R}\right)^4 -
\frac{9}{8} \left(\frac{\varpi}{R}\right)^2 \left(\frac{z}{R}\right)^2
+ \frac{3}{8} \left(\frac{z}{R}\right)^4 + {\cal O}(6) \right)
\label{poteq}
\end{eqnarray}
for the axisymmetric potential corresponding to the tidal field of
M31, up to fourth order in $z/R$ and $\varpi/R$, with $M(R)$ the mass
of M31 enclosed within a sphere of radius $R$ \cite{kzs02}.

We plot the gravitational potential of the best fit dynamical model
for NGC205 as a function of radius in the equatorial plane. The top
curve is the potential of the model in the absence of M31. From top to
bottom, the other curves indicate the potential in the presence of the
axisymmetrized tidal field of M31 for relative distances of 40, 20, 10
and 5~kpc between NGC205 and M31. Even for the minimum current
relative distance allowed by the observations, i.e. 10~kpc, the tidal
influence of the giant spiral galaxy is negligible within the region
for which we have kinematical data, that is within the inner
1.2~kpc. In that case, stars beyond 2.8~kpc (700{\arcsec}) are
unbound. If NGC205 has passed by M31 with a pericentric distance as
small as 5~kpc, stars are unbound beyond a radius of $\sim
1.5$~kpc. These values can be compared with the analytical estimate
\begin{equation}
r_{\rm tidal} \simeq r_{\rm peri} \left( \frac{M_{\rm sat}}{x\, M_{\rm
gal}} \right)^{1/3},
\end{equation}
with $r_{\rm peri}$ the orbital radius of a satellite galaxy with mass
$M_{\rm sat}$ around a galaxy with mass $M_{\rm gal}$, and $x=9$ or 1,
for stars inside the satellite galaxy on prograde or retrograde
circular orbits, respectively \cite{re06}. For a pericentric distance
of 10 kpc, this yields $r_{\rm tidal} = 1.0$ to 2.0~kpc (depending on
$x$); for a pericentric distance of 5 kpc, this yields $r_{\rm tidal}
= 0.6$ to 1.3 kpc. We substituted the mass of M31 enclosed within a
radius $\rm r_{\rm peri}$ for $M_{\rm gal}$. These are also the radii
where in Fig. \ref{pot205} the "effective" potential starts to deviate
significantly from that of NGC205 alone. This makes us confident that
we are providing an adequate description of the tidal enfluence of M31
on NGC205.

Small pericentric distances, i.e. smaller than $\sim 5-10$~kpc, which
may have occurred in the past, lead to a large effect on the force
field within the inner 1.2~kpc but also give rise to a tidal radius
well within the region constrained by the photometric and kinematical
data, in which case the construnction of equilibrium models itself
becomes meaningless. In short, our results are robust if NGC205 has
kept a distance of at least $\sim 10$~kpc from M31.

\subsection{NGC147}

\subsubsection{New and previous kinematical work}

\begin{figure}
\vspace*{14cm}
\special{hscale=90 vscale=90 hsize=500 vsize=500
hoffset=-15 voffset=-190 angle=0 psfile="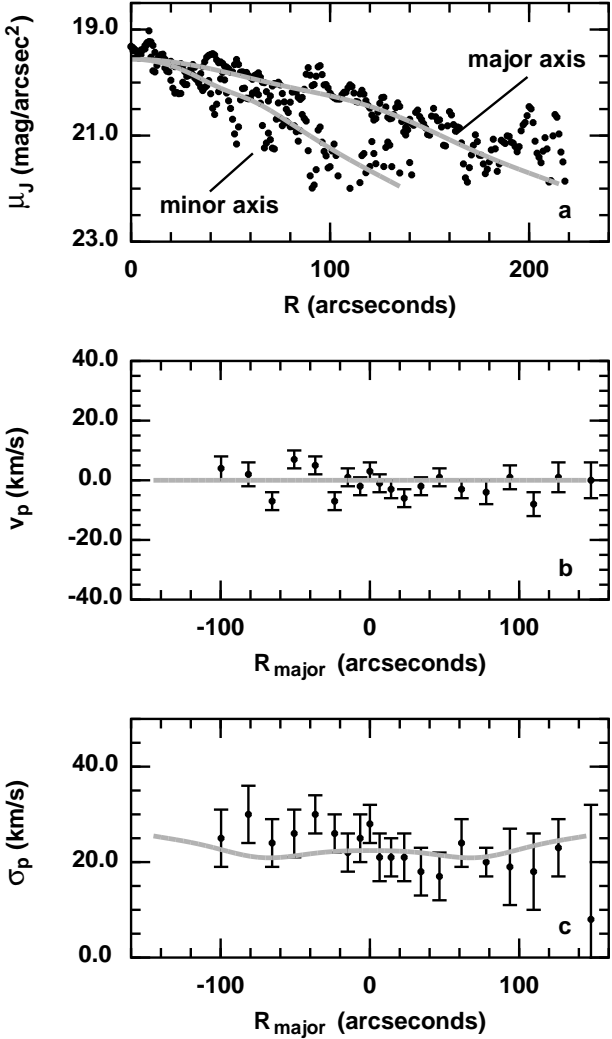"}
\caption{Fit to the kinematics of NGC147. Panel {\bf a}:~the J-band
surface brightness, $\mu_J$, along major and minor axis; panel {\bf
b}:~the mean streaming velocity along the major axis, $v_p$; panel
{\bf c}:~the major-axis velocity dispersion, $\sigma_p$. The black
dots are the data points, the grey curves correspond to the best
fitting dynamical model. The model was fitted to the surface
brightness on a grid covering the whole face of the galaxy but for
clarity we only plot the major and minor axis profiles in this
figure. The major-axis velocity dispersion remains approximately
constant at $\sigma_p \sim 25$~km~s$^{-1}$ out to the last data point,
at 2.3{\arcmin}. \label{kin147}}
\end{figure}

The major and minor axis J-band photometry and the major-axis
kinematics of NGC147 as measured by Simien \& Prugniel
\shortcite{sp02}, including the additional exposures obtained in
January 2002 and 2003, are presented in Fig. \ref{kin147}. The
kinematics of NGC147 extend out to $\sim 150''$ (0.5~kpc or
0.8~$R_{\rm e}$). The projected velocity dispersion is roughly
constant, at about $\sigma_p \approx 25$~km~s$^{-1}$. The rotation
curve is consistent with NGC147 having zero rotation. The velocity
dispersion profile is in good agreement with the one presented by
Bender et al. \shortcite{be91}. However, these authors found
significant rotation in NGC147, with an amplitude of $\sim
10$~km~s$^{-1}$ at a radius of 100{\arcsec}. They estimated the B-band
mass-to-light ratio of NGC147 at $(M/L)_{\rm B} = 7 \pm 3
\,M_\odot/L_{\odot,\rm B}$.

\subsubsection{The dynamical mass and mass-to-light ratio}

The total mass within a 2~$R_{\rm e}$ radius sphere is estimated at
$M(2\,R_{\rm e}) = 3.0^{+2.4}_{-1.8} \times 10^8 \,M_\odot$,
corresponding to a J-band mass-to-light ratio within the inner
2~$R_{\rm e}$ of $(M/L)_{\rm J}
=3.4^{+2.7}_{-2.0}\,M_\odot/L_{\odot,\rm J}$ or a B-band mass-to-light
ratio of $(M/L)_{\rm B} =4.0^{+3.2}_{-2.4}\,M_\odot/L_{\odot,\rm
B}$. The mean metallicity of the stellar population is [Fe/H$]\approx
-1.0$ \cite{ha97,mc05}. There is no evidence that NGC147 contains a
population of intermediate-age stars \cite{r98}; Mould, Kristian, and
Da Costa \shortcite{mkd83} estimate that 90\% of the stellar
population is older than 12~Gyr. Using 12~Gyr as a rough estimate for
the mean age of the stellar population, the stellar mass-to-light
ratio is $(M/L)_{\rm J} \sim 1.6\,M_\odot/L_{\odot,\rm J}$ or
$(M/L)_{\rm B} \sim 2.6\,M_\odot/L_{\odot,\rm B}$
\cite{wo94}. Comparing this with the dynamical mass-to-light ratio of
the best fit dynamical model, we find that NGC147 consists of about
50\% luminous matter, by mass, and of 50\% dark matter.

\begin{figure}
\vspace*{4.5cm}
\special{hscale=90 vscale=90 hsize=500 vsize=500
hoffset=-15 voffset=-190 angle=0 psfile="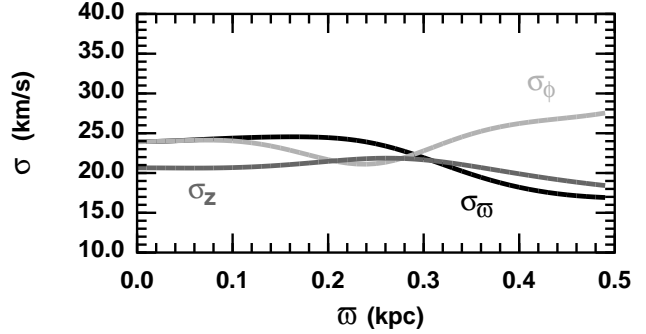"}
\caption{The radial, tangential, and vertical components of the
velocity dispersion tensor of the best fit model of NGC147. All
quantities are plotted as a function of radius within the equatorial
plane. The spatial stellar density profile within the equatorial plane
is consistent with being exponentially declining from a central
density of $0.32 \,L_{\rm J,\odot}\,$pc$^{-3}$ with an exponential
scale-length of 0.31~kpc.
\label{spc147}}
\end{figure}

\subsubsection{The internal dynamics}

The stellar density and the radial, tangential, and vertical
components of the velocity dispersion tensor of the best fit model for
NGC147 are presented in Fig. \ref{spc147}. The velocity dispersion
tensor is almost isotropic, with $\sigma \approx 20-25$~km~s$^{-1}$,
up to a radius of 0.3~kpc, with the vertical velocity dispersion being
somewhat lower to help flatten the galaxy. Outside 0.3~kpc, the
tangential velocity dispersion rises up to $\sigma_\phi \approx
27$~km~s$^{-1}$ in sync with the projected velocity dispersion
(Fig. \ref{kin147}).

\subsection{NGC185}

\begin{figure}
\vspace*{14cm}
\special{hscale=90 vscale=90 hsize=500 vsize=400
hoffset=-30 voffset=-200 angle=0 psfile="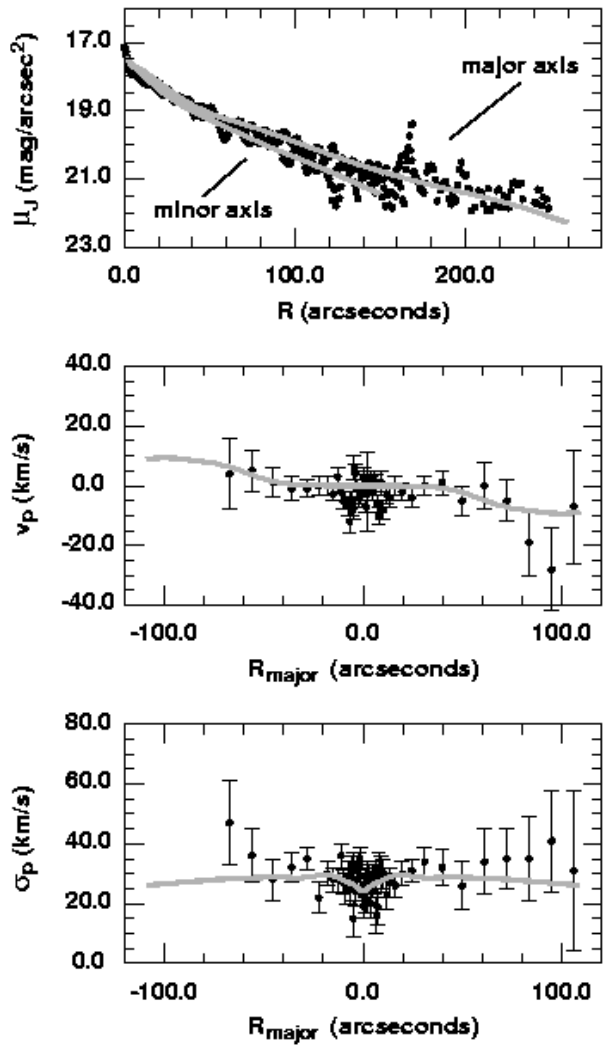"}
\caption{Fit to the streaming velocity along the major axis, $v_p$,
the major and minor axis velocity dispersion, $\sigma_p$, and the
J-band surface brightness, $\mu_J$, of NGC185 with a 3I dynamical
model with inclination $i=50^\circ$. The black dots are the data
points, the grey curves correspond to the model. The model was fitted
to the surface brightness on a grid covering the whole face of the
galaxy but for clarity we only plot the major and minor axis profiles
in this figure. \label{kin185}}
\end{figure}


\subsubsection{New and previous kinematical work}

The major and minor axis J-band photometry and major-axis kinematics
of NGC185 as measured by Simien \& Prugniel \shortcite{sp02} are
presented in Fig. \ref{kin185}. The kinematics of NGC185 extend out to
$\sim 100${\arcsec} (0.3~kpc or 1.1~$R_{\rm e}$). The major-axis
velocity dispersion is roughly constant at about 30~km~s$^{-1}$, in
good agreement with Bender et al. \shortcite{be91}. There is a hint of
rotation, at variance with Bender et al. \shortcite{be91}, but, as in
the case of NGC147, sky subtraction at these low surface-brightness
levels is very difficult and this might affect the
measurements. Bender et al. \shortcite{be91} estimate the B-band
mass-to-light ratio of NGC185 at $(M/L)_{\rm B} = 5 \pm 2
\,M_\odot/L_{\odot,\rm B}$. Held et al. \shortcite{he92} found an
essentially flat velocity dispersion profile, with $\sigma \approx
30$~km/s. Interestingly, their major-axis velocity dispersion profile
shows a central peak, with the dispersion rising to a central value of
50~km/s. These authors used a spherically symmetric, isotropic model
with an exponentially declining density profile to estimate the
central mass-to-light ratio of NGC185 at $(M/L)_{\rm B} \sim
3\,M_\odot/L_{\odot,\rm B}$.

\subsubsection{The dynamical mass and mass-to-light ratio}

The best fitting model, with $i=50^{\circ}$ and an intrinsic axial
ratio $q_{\rm intr}=0.62$, yields our best estimate for the
mass of NGC185. The error-bar on this quantity is estimated taking
into account all inclination angles. We estimate the mass within a
2~$R_{\rm e}$ radius sphere at $M(2\,R_{\rm e}) = 2.6^{+0.8}_{-0.6}
\times 10^8 \,M_\odot$, corresponding to a J-band mass-to-light ratio
within the inner 2~$R_{\rm e}$ of $(M/L)_{\rm J}
=2.2^{+0.6}_{-0.5}\,M_\odot/L_{\odot,\rm J}$ or a B-band mass-to-light
ratio of $(M/L)_{\rm B} = 3.0^{+1.0}_{-0.7}\,M_\odot/L_{\odot,\rm
B}$. This result is in good agreement with the one obtained by Held et
al. \shortcite{he92}. The mean metallicity of the stellar population
is estimated at [Fe/H$]=-1.43 \pm 0.15$ by Mart\'{\i}nez-Delgado \&
D. \& Aparicio \shortcite{ma98}. Lee et al. \shortcite{l1} find NGC185
to contain a complex stellar population with a mean metallicity of
[Fe/H$]=- 1.23 \pm 0.16$. McConnachie~et al. \shortcite{mc05} derive a
value [Fe/H$]=-1.2$. Using 10~Gyr as a rough estimate for the mean age
of the stellar population \cite{mag99}, the stellar mass-to-light
ratio is $(M/L)_{\rm J} \sim 1.5 \,M_\odot/L_{\odot,\rm J}$ or
$(M/L)_{\rm B} \sim 2.1\,M_\odot/L_{\odot,\rm B}$ \cite{wo94}. In
other words, NGC185 consists of about 60\% luminous matter, by mass,
and of 40\% dark matter.

\subsubsection{The internal dynamics}

The stellar density and the radial, tangential, and vertical
components of the velocity dispersion tensor of the best fit models
for NGC185 are presented in Fig. 13. The velocity
dispersion tensor of the edge-on model ($i=90^{\circ}$, see left
column of Fig. 13) has a remarkably large vertical velocity
dispersion, peaking up to $\sigma_z \sim 40$~km~s$^{-1}$, in the inner
parts of the galaxy. This feature is not just an oddity of this one
dynamical model but is common to all edge-on models we fitted to
NGC185. Since within the context of axisymmetric models, viewed
edge-on, the vertical velocity dispersion is determined almost solely
by the photometry, we have to conclude that if NGC185 is viewed
edge-on, it can only maintain its round shape by having the vertical
velocity dispersion much larger than the radial velocity dispersion,
which is set by the observed major-axis velocity dispersion. The model
with $i=50^{\circ}$ and an intrinsic axial ratio $q_{\rm intr}=0.62$,
is the most nearly isotropic of all models we constructed for NGC185,
see the middle column of Fig. 13. The rotation velocity
of the stars flattens at about 22~km~s$^{-1}$ in this model, versus a
rotation velocity of 8~km~s$^{-1}$ in the edge-on model. Since $(i)$
the intrinsic axial ratio of this model is still fairly close to the
peak of the intrinsic shape distribution of dEs, and $(ii)$, like the
best fitting models for NGC147 and NGC205, it is very nearly
isotropic, this is our preferred best model for NGC185. The
intrinsically very flattened model, with $q_{\rm intr}=0.5$ and
$i=45^{\circ}$, see the right column of Fig. 12, has a much larger
rotation velocity, its peak projected rotation velocity is somewhat
higher than that of the $i=90^{\circ}$ model but still safely within
the error-bars. The vertical velocity dispersion $\sigma_z$ is now
much smaller than in the less inclined models. Although this model
agrees with the observations to within the (quite substantial)
error-bars, it fails to reproduce the outward rise of the major-axis
velocity dispersion. Hence, with better data it should be possible to
better constrain the inclination of NGC185 on purely dynamical
grounds.

\section{Conclusions} \label{conc}

\begin{figure*}
 \vbox to220mm{\vfil 
\vspace*{24.5cm}
\special{hscale=100 vscale=100 hsize=500 vsize=700
hoffset=-60 voffset=-40 angle=0 psfile="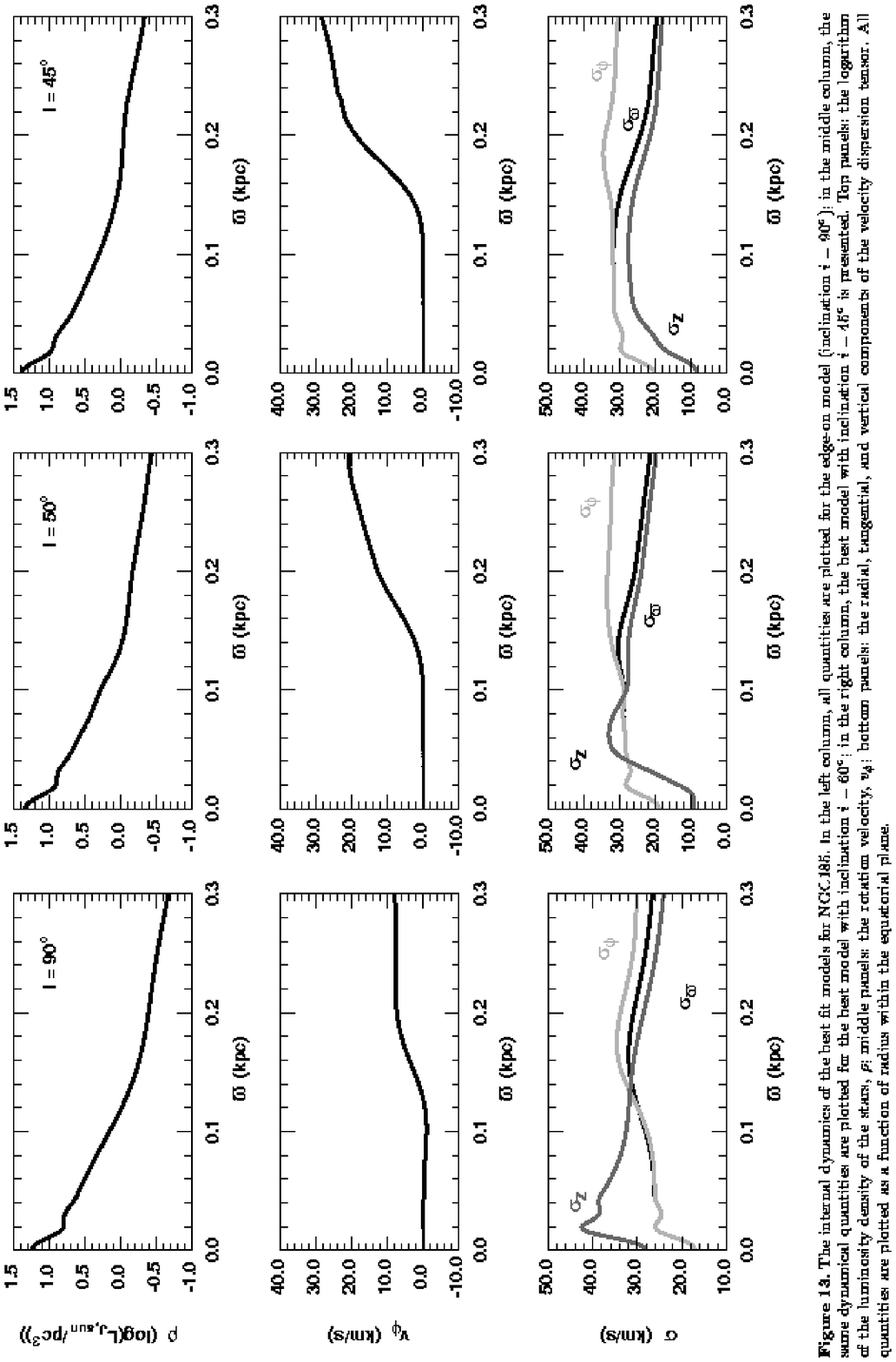"}
\vfil}
\label{spc185_90}
\end{figure*}

In this paper, we present dynamical models for NGC147, NGC185, and
NGC205. NGC205 and NGC147 have already a flattened apparent shape so
we modeled them assuming them to be viewed edge-on
($i=90^\circ$). NGC185, on the other hand, has a much rounder
projected shape than the two other Local Group dEs, so we produced
models for it assuming inclination angles $i=90^\circ,\, 60^\circ,\,
55^\circ,\, 50^\circ$, and $45^\circ$.

From the range of models that is compatible with the data at the 95\%
confidence level, we estimate the mass of NGC205 within a 2~$R_{\rm
e}$ radius sphere at $M(2\,R_{\rm e}) = 10.2^{+3.3}_{-2.0} \times 10^8
\,M_\odot$, corresponding to a J-band mass-to-light ratio within the
inner 2~$R_{\rm e}$ of $(M/L)_{\rm J}
=2.2^{+0.7}_{-0.4}\,M_\odot/L_{\odot,\rm J}$ or a B-band mass-to-light
ratio of $(M/L)_{\rm B} = 4.5^{+1.5}_{-0.9}\,M_\odot/L_{\odot,\rm
B}$. The best fitting 3I model for NGC205 very accurately reproduces
the photometric and kinematical data, such as the very bright nucleus
and the central drop in the velocity dispersion. Overall, the velocity
dispersion tensor is quite isotropic. Only beyond a radius of 0.8~kpc
does the tangential anisotropy increase to keep up with the observed
rise of the projected velocity dispersion along the major axis. The
very bright nucleus shows up in the model as spherically symmetric,
dynamically very cold substructure, which can be modeled as an
isotropic star cluster with a total mass of $1.36 \times
10^6\,M_\odot$, a one-dimensional velocity dispersion $\sigma \approx
7$~km~s$^{-1}$ and a truncation radius of about 50~pc. Thus, the
nucleus of NGC205 is structurally and dynamically not unlike a massive
globular cluster.

We estimate the total mass of NGC147 within a 2~$R_{\rm e}$ radius
sphere at $M(2\,R_{\rm e}) = 3.0^{+2.4}_{-1.8} \times 10^8 \,M_\odot$,
corresponding to a J-band mass-to-light ratio within the inner
2~$R_{\rm e}$ of $(M/L)_{\rm J}
=3.4^{+2.7}_{-2.0}\,M_\odot/L_{\odot,\rm J}$ or a B-band mass-to-light
ratio of $(M/L)_{\rm B} =4.0^{+3.2}_{-2.4}\,M_\odot/L_{\odot,\rm
B}$. As in the case of NGC205, the velocity dispersion tensor of the
best fit model for NGC147 is remarkably isotropic. 

We estimate the total mass of NGC185 within a 2~$R_{\rm e}$ radius
sphere at $M(2\,R_{\rm e}) = 2.6^{+0.8}_{-0.6} \times 10^8 \,M_\odot$,
corresponding to a J-band mass-to-light ratio within the inner
2~$R_{\rm e}$ of $(M/L)_{\rm J}
=2.2^{+0.6}_{-0.5}\,M_\odot/L_{\odot,\rm J}$ or a B-band mass-to-light
ratio of $(M/L)_{\rm B} = 3.0^{+1.0}_{-0.7}\,M_\odot/L_{\odot,\rm
B}$. The model with $i=50^{\circ}$ has an intrinsic flattening that is
still close to the peak of the intrinsic shape distribution of dEs and
it, like the best fitting models for NGC147 and NGC205, is nearly
isotropic. Therefore, this is our preferred model for NGC185.

Summarizing, our dynamical models for the three Local Group dEs show
them to have nearly isotropic velocity dispersion tensors. They have
very similar mass-to-light ratios, of the order of $(M/L)_{\rm B} \sim
4 \,M_\odot/L_{\odot,\rm B}$. While this is still larger than the
expected mass-to-light ratio of the stellar populations of these
galaxies, it is about a factor of two smaller than the mass-to-light
ratios derived by many previous studies. This rather large difference
is due in part to differences in the data analysis. Depending on the
spectral resolution, the signal-to-noise ratio of the spectra, and the
method of extracting kinematical data from the spectra, the velocity
dispersion estimates for the same galaxy can differ by a factor of
$1.5-2$ between different authors, with lower resolution observations
tending to yield higher values for the velocity dispersion. This,
exacerbated by the difficulty of measuring core-radii and surface
brightnesses of such diffuse objects, required for use in the standard
King formula \cite{rt86}, seems to account for at least part of the
large spread on the $M/L$ estimates. Therefore, these dwarf galaxies
contain much less dark matter than was believed to be the case up to
now. Still, within the inner two half-light radii, about $40-50$\% of
the mass is in the form of dark matter, so dEs contain as much dark
matter as bright ellipticals. Indeed, Gerhard et al. \shortcite{ge01}
produced dynamical models for a sample of 21 luminous ellipticals and
found that the dark matter contributes $10-40$\% of the mass within
$1\,R_{\rm e}$ and equal interior mass of dark and luminous matter
occurs at $2-4\,R_{\rm e}$.

\section*{Acknowledgments}
This research has made use of the NASA/IPAC Infrared Science Archive,
which is operated by the Jet Propulsion Laboratory, California
Institute of Technology, under contract with the National Aeronautics
and Space Administration. We thank the referee for his constructive
remarks.

\appendix

\section{The three-integral components} \label{app}

For a given potential, we wish to find the DF that best reproduces the
kinematical information. As a first step, the DF is written as a
weighted sum of basis functions, called ``components''. Here, we use
components of the form
\begin{eqnarray}
F^{i,\epsilon_i}(E,I_2,I_3) &=& (E - E_{0,i})^{\sigma_i} (I_2 -
I_{0,i})^{\tau_i} I_3^{\rho_i}, \nonumber \\ && \hspace{3.93em}\,{\rm
if}\,\,E>E_{0,i},\,\epsilon_i I_2 > I_{0,i} \ge 0 \nonumber \\ &=& 0,
\hspace{3em}\,{\rm if}\,E \le E_{0,i}\,\,{\rm or}\,\,\epsilon_i I_2 \le I_{0,i}
\end{eqnarray}
with $\sigma_i$, $\tau_i$, and $\rho_i$ integer numbers. $E_{0,i}$ is
a lower bound on the binding energy, which defines the outer boundary
within which the component is non-zero. The condition $\epsilon_i I_2
> I_{0,i} \ge 0$, with $\epsilon_i \in [-1,+1]$ defines the
``handedness'' of the component. If $\epsilon_i=+1$, only orbits with
positive angular momentum $I_2 > I_0$ are populated and we call the
component $F^{i,+}$ ``right-handed''. If $\epsilon_i=-1$, only orbits
with negative angular momentum $I_2 < -I_0$ are populated and the
component $F^{i,-}$ is ``left-handed''. Components that populate
prograde and retrograde orbits equally, labeled with $\epsilon_i=0$,
are constructed as $F^{i,0}(E,I_2,I_3) = F^{i,+}(E,I_2,I_3) +
F^{i,-}(E,I_2,I_3)$. Components with $I_0 \ne 0$ and $E_0 \ne 0$
populate orbits within a torus-shaped region of space.

The spatial velocity moments, $\mu^{i,+}_{2l,m,2n}(\varpi,z)$, of the
component $F^{i,+}(E,I_2,I_3)$ are given by:
\begin{eqnarray}
\lefteqn{\mu^{i,+}_{2l,m,2n}(\varpi,z)} \nonumber \\ &=& \int
F^{i,+}(E,I_2,I_3) v_\lambda^{2l} v_\phi^m v_\nu^{2n} \,d v_\lambda
v_\phi v_\nu \\ &=& \frac{2^{l+n+(m+1)/2}}{\varpi^{m+1}
(\lambda-\nu)^{l+n}} \int_{E_{0,i}}^\psi (E-E_{0,i})^{\sigma_i} dE
\nonumber \\
&& \times \, 
\int_{I_{0,i}}^{\varpi^2(\psi-E)} I_2^{(m-1)/2} (I_2-I_{0,i})^{\tau_i}
dI_2 \nonumber \\
&& \times \, \int_{I_3^-}^{I_3^+} I_3^{\rho_i} (I_3^+ - I_3)^{l-1/2} (I_3 -
I_3^-)^{n-1/2} dI_3,
\end{eqnarray}
with
\begin{eqnarray}
I_3^+ &=& (\lambda+\gamma)\left(G(\lambda)-E -
\frac{I_2}{\lambda+\alpha}\right), \\ I_3^- &=&
(\nu+\gamma)\left(G(\nu)-E-\frac{I_2}{\nu+\alpha} \right).
\end{eqnarray}
This integration can be performed analytically by repeatedly using the
lemma
\begin{eqnarray}
\lefteqn{\int_a^b x^i (b-x)^j (x-a)^k dx = B(k+1,j+1) } \nonumber \\
&& \times \, (b-a)^{j+k+1} b^i
\,_2F_1\left(-i,j+1;j+k+2;1-\frac{b}{a}\right),
\end{eqnarray}
with $B$ the Euler beta function and $_2F_1$ the hypergeometric
function. Thus, one obtains the following expression for the spatial
velocity moments:
\begin{eqnarray}
 \lefteqn{\mu^{i,+}_{2l,m,2n}(\varpi,z)} \nonumber\\ &=& {\cal
F}^{\,i}_{\,l,m,n} \varpi^{2 \tau_i} \left( \psi - E_{0,i} -
\frac{I_{0,i}}{\varpi^2} \right)^{\sigma_i+\tau_i+n+l+2}\nonumber \\
&\times& \!\!  (\psi-E_{0,i})^{(m-1)/2} \left[ (\lambda+\gamma)
(G(\lambda)-E_{0,i})\right]^\rho_i \nonumber \\ &\times& \!\!
\sum_{r=0}^{\rho_i} \frac{(-\rho_i)_r (l+1/2)_r}{r!  (\sigma_i+\tau_i
+ n + l + 3)_r} \nonumber \\ && \hspace{6em} \left(
\frac{(\lambda-\nu)(\psi-E_{0,i}-I_{0,i}/\varpi^2)}{(\lambda+\gamma)(G(\lambda)-E_{0,i})}
\right)^r \nonumber \\ &\times& \!\! \sum_{s=0}^{\rho_i-r}
\frac{(r-\rho_i)_s (\sigma_i+1)_s}{s! (\sigma_i+\tau_i+n+l+r+3)_s}
\left( \frac{\psi-E_{0,i}-I_{0,i}/\varpi^2}{G(\lambda)-E_{0,i}}
\right)^s \nonumber \\ &\times& \!\! \sum_{p=0}^{\rho_i-r-s}
\frac{(r+s-\rho_i)_p}{p!}  \left(
\frac{(\alpha+\nu)(\psi-E_{0,i})}{(\alpha-\gamma)
(G(\lambda)-E_{0,i})} \right)^p \nonumber \\ &\times&\!\! _2F_1\left(
-p - \frac{m-1}{2}, \sigma_i+\tau_i+n+l+r+s+2; \right. \nonumber \\&&
\hspace{1em} \left. \sigma_i+\tau_i+n+l+r+s+3;1 -
\frac{I_{0,i}}{\varpi^2(\psi-E_{0,i})} \right).
\end{eqnarray}
with the forefactor ${\cal F}^{\,i}_{\,l,m,n}$ given by
\begin{eqnarray}
\lefteqn{ {\cal F}^{\,i}_{\,l,m,n} =} \nonumber \\ && 2^{n+l+(m+1)/2} \frac{\Gamma(\sigma_i+1) \Gamma(\tau_i+1)
\Gamma(n+\frac{1}{2})
\Gamma(l+\frac{1}{2})}{\Gamma(\sigma_i+\tau_i+n+l+3)}.
\end{eqnarray}
The velocity moments of the left-handed component
$F^{i,-}(E,I_2,I_3)$, $\mu^{i,-}_{2l,m,2n}$, are linked to those of
its right-handed analog $F^{i,+}(E,I_2,I_3)$ via the relation
$\mu^{i,-}_{2l,m,2n} = (-1)^m \mu^{i,+}_{2l,m,2n}$. The spatial
velocity moments with respect to the cylindrical velocity components
$v_\varpi$, $v_\phi$, and $v_z$ can be constructed from the moments
$\mu^{i,\epsilon_i}_{2l,m,2n}$ using the transformation
\begin{equation}
\left( 
\begin{array}{c}
v_\varpi \\
{\rm sign} (z) \,v_z
\end{array}
\right) = \left(
		\begin{array}{cc}
		\cos \Theta & -\sin \Theta \\
		\sin \Theta & \cos \Theta
		\end{array}
		\right)
\left(
		\begin{array}{c}
		v_\lambda \\
		v_\nu
		\end{array}
		\right)
\end{equation}
with
\begin{equation}
\cos \Theta = \sqrt{
\frac{(\nu+\alpha)(\lambda+\gamma)}{(\alpha-\gamma)(\lambda-\nu)}}, \,
\sin \Theta = \sqrt{
\frac{(\lambda+\alpha)(\nu+\gamma)}{(\gamma-\alpha)(\lambda-\nu)}}.
\end{equation}

\bsp \label{lastpage} 
\begin{thebibliography}{99}
\bibitem[\protect\citename{Baade }1944a]{b44a} Baade, W., 1944, ApJ, 100,
137
\bibitem[\protect\citename{Baade }1944b]{b44b} Baade, W., 1944, ApJ,
100, 147
\bibitem[\protect\citename{Bender et al. }1991]{be91} Bender, R.,
Paquet, A., Nieto, J.-L., 1991, A\&A, 246, 349
\bibitem[\protect\citename{Binggeli \& Popescu }1995]{bp95} Binggeli,
B. \& Popescu, C., 1995, A\&A, 298, 63
\bibitem[\protect\citename{Butler \& Mart\'{\i}nez-Delgado
}2005]{bm05} Butler, D. J. \& Mart\'{\i}nez-Delgado, D., 2005, AJ,
129, 2217
\bibitem[\protect\citename{Carter \& Sadler }1990]{cs90} Carter, D. \&
Sadler, E. M., 1990, MNRAS, 245P, 12
\bibitem[\protect\citename{Choi et al. }2002]{ch02} Choi, P. I.,
Guhathakurta, P., Johnston, K., 2002, AJ, 124, 310
\bibitem[\protect\citename{De Bruyne et al. }2001]{deb01} De Bruyne,
V., Dejonghe, H., Pizzella, A., Bernardi, M., Zeilinger, W. W., 2001,
ApJ, 546, 903
\bibitem[\protect\citename{Dejonghe~\&~de Zeeuw }1988]{dz88} Dejonghe,
H., \& de Zeeuw, T., 1988, ApJ, 329, 720
\bibitem[\protect\citename{Dejonghe }1989]{d89} Dejonghe, H., 1989,
ApJ, 343, 113
\bibitem[\protect\citename{Dejonghe \& Merritt }1992]{dm92} Dejonghe,
H. \& Merritt, D., 1992, ApJ, 391, 531
\bibitem[\protect\citename{Dejonghe~et al. }1996]{de96} Dejonghe, H.,
De Bruyne, V., Vauterin, P., Zeilinger, W. W., 1996, A\&A, 306, 363
\bibitem[\protect\citename{de Vaucouleurs et al. }1991]{rc3} de
Vaucouleurs, G., de Vaucouleurs, A., Corwin JR., H. G., Buta,
R. J. Paturel, G., Fouque, P., 1991, Third reference catalogue of
bright galaxies, version 3.9
\bibitem[\protect\citename{Evans \& Collett }1994]{ec94} Evans, N.~W.,
\& Collett, J.~L., 1994, ApJ, 420, L67
\bibitem[\protect\citename{Ferguson~\&~Binggeli }1994]{fb94} Ferguson,
H.~C., \& Binggeli, B. 1994, A\&ARv, 6, 67
\bibitem[\protect\citename{Franx et al. }1989]{f89} Franx, M.,
Illingworth, G., Heckman, T., 1989, ApJ, 344, 613
\bibitem[\protect\citename{Geha, Guhathakurta, van der Marel
}2002]{g02} Geha, M., Guhathakurta, P., van der Marel, R. P., 2002,
AJ, 124, 3073
\bibitem[\protect\citename{Geha et al. }2006]{g04} Geha, M.,
Guhathakurta, P., Rich, R. M., Cooper, M. C., 2006, AJ, 131, 332
\bibitem[\protect\citename{Gerhard \& Binney }1996]{gb96} Gerhard,
O. \& Binney, J., 1996, MNRAS, 279, 993
\bibitem[\protect\citename{Gerhard et al. }2001]{ge01} Gerhard, O.,
Kronawitter, A., Saglia, R. P., Bender, R., 2001, AJ, 121, 1936
\bibitem[\protect\citename{Han et al. }1997]{ha97} Han, M., Hoessel,
J. G., Gallagher, J. S., {\sc iii}, Hotsman, J., Stetson, P. B. et
al., 1997, AJ, 113, 1001
\bibitem[\protect\citename{Held et al. }1990]{he90} Held, E. V.,
Mould, J. R., de Zeeuw, P. T., 1990, AJ, 100, 415
\bibitem[\protect\citename{Held et al. }1992]{he92} Held, E. V., de
Zeeuw, T., Mould, J., Picard, A., 1992, AJ, 103, 851
\bibitem[\protect\citename{Hodge }1973]{h73} Hodge, P. W., 1973, ApJ,
182, 67
\bibitem[\protect\citename{Jarrett et al. }2003]{ja03} Jarrett, T. H.,
Chester, T., Cutri, R., Schneider, S. E., Huchra, J. P., 2003, AJ,
125, 525
\bibitem[\protect\citename{Jones et al. }1996]{jo96} Jones, D. H.,
Mould, J. R., Watson, A. M., Grillmair, C., Gallagher {\sc iii},
J. S., Ballester, G. E., Burrows C. J., Casertano, S., et al., 1996,
ApJ, 466, 742
\bibitem[\protect\citename{Kim \& Lee }1998]{kl98} Kim, S. C. \& Lee,
M. G., 1998, JKAS, 31, 51
\bibitem[\protect\citename{Klypin, Zhao, Somerville }2002]{kzs02}
Klypin, A., Zhao, H., Somerville, R. S., 2002, ApJ, 573, 597
\bibitem[\protect\citename{Koleva et al. }2006]{ko06} Koleva, M.,
Bavouzet, N., Chilingarian, I., Prugniel, P., 2006, in "Scientific
prospectives for 3D spectroscopy", Kissler-Patig, M. and Walsh,
J. (Edts), in press.
\bibitem[\protect\citename{Lee et al. }1993]{l1} Lee, M. G., Freedman,
W. L., Madore, B. F., 1993, AJ, 106, 964
\bibitem[\protect\citename{Lee }1996]{l2} Lee, M. G., 1996, AJ, 112,
1438
\bibitem[\protect\citename{Lotz et al. }2001]{lo01} Lotz, J. M.,
Telford, R., Ferguson, H. C., Miller, B. W., Stiavelli, M., Mack, J.,
ApJ, 552, 572
\bibitem[\protect\citename{Mart\'{\i}nez-Delgado, Aparicio, Gallart
}1999]{mag99} Mart\'{\i}nez-Delgado, D., Aparicio, A., Gallart, C.,
1999, ApJ, 118, 2229
\bibitem[\protect\citename{Mart\'{\i}nez-Delgado \& Aparicio
}1998]{ma98} Mart\'{\i}nez-Delgado, D. \& Aparicio, A., AJ, 115, 1462
\bibitem[\protect\citename{McConnachie~et al. }2005]{mc05}
McConnachie, A. W., Irwin, M. J., Ferguson, A. M. N., Ibata, R. A.,
Lewis, G. F., Tanvir, N., 2005, MNRAS, 356, 979
\bibitem[\protect\citename{Merritt \& Quinlan }1998]{mq98} Merritt,
D., \& Quinlan, G.~D., 1998, ApJ, 498, 625
\bibitem[\protect\citename{Messier }1801]{m01} Messier, C., 1798,
Observations Astronomiques, 1770-1774, Connaissance des Temps pour
l'an {\sc ix}, 461
\bibitem[\protect\citename{Mould, Kristian, Da Costa }1983]{mkd83}
Mould, J. R., Kristian, J., Da Costa, G. S., 1983, ApJ, 270, 471
\bibitem[\protect\citename{Peletier }1993]{p93} Peletier, R. F., 1993,
A\&A, 271, 51
\bibitem[\protect\citename{Prugniel \& Soubiran }2001]{ps01} Prugniel,
P. \& Soubiran, C., 2001, A\&A, 369, 1048
\bibitem[\protect\citename{Read et al. }2006]{re06} Read, J. I.,
Wilkinson, M. I., Evans, N. W., Gilmore, G., Kleyna, J. T., 2006,
MNRAS, 366, 429
\bibitem[\protect\citename{Renzini }1998]{r98} Renzini, A., 1993, AJ,
115, 2459
\bibitem[\protect\citename{Richer et al. }1998]{ri98} Richer, M. G.,
McCall, M. L., Stasi\'nska, G., 1998, A\&A, 340, 67
\bibitem[\protect\citename{Richstone \& Tremaine }1986]{rt86}
Richstone, D. O. \& Tremaine, S., 1986, AJ, 92, 72
\bibitem[\protect\citename{Simien \& Prugniel }2002]{sp02} Simien,
F. \& Prugniel, Ph., 2002, A\&A, 384, 371
\bibitem[\protect\citename{Tremaine, Ostriker, Spitzer }1975]{tos75}
Tremaine, S. D., Ostriker, J. P., Spitzer, L., 1975, ApJ, 196, 407
\bibitem[\protect\citename{Valluri et al. }2005]{v05} Valluri, M.,
Ferrarese, L., Merritt, D., Joseph, C. L., 2005, ApJ, 628, 137
\bibitem[\protect\citename{van den Bergh }1998]{s98} van den Bergh,
S., 1998, AJ, 116, 1688
\bibitem[\protect\citename{Worthey }1994]{wo94} Worthey, G., 1994,
ApJS, 95, 107
\end{thebibliography}
\end{document}